\documentclass[12pt,draftclsnofoot,onecolumn]{IEEEtran}

\usepackage[dvips]{graphicx}
\usepackage{amssymb}
\usepackage[cmex10]{amsmath}
\usepackage[caption=false]{subfig}
\usepackage{multirow}
\usepackage{array}

\newtheorem{definition}{Definition}
\newtheorem{proposition}{Proposition}
\newtheorem{corollary}{Corollary}
\newtheorem{assumption}{Assumption}
\newtheorem{lemma}{Lemma}

\begin{document}

\title{The Theory of Intervention Games for Resource Sharing in Wireless Communications}
\author{Jaeok Park and Mihaela van der Schaar
\thanks{A preliminary version of this paper was presented in part at the 30th IEEE International Conference on Computer
Communications (IEEE INFOCOM 2011), Shanghai, China, April 10--15, 2011 \cite{ParkInfocom}.
The authors are with Electrical Engineering
Department, University of California, Los Angeles (UCLA), 420
Westwood Plaza, Los Angeles, CA 90095-1594, USA. e-mail: \{jaeok,
mihaela\}@ee.ucla.edu.}}
\date{}

\maketitle


\begin{abstract}
This paper develops a game-theoretic framework for the design and analysis of a new class of
incentive schemes called intervention schemes. We formulate intervention games, propose
a solution concept of intervention equilibrium, and prove its existence in a finite intervention game.
We apply our framework to resource sharing scenarios in wireless communications, whose
non-cooperative outcomes without intervention yield suboptimal performance.
We derive analytical results and analyze illustrative examples in the cases
of imperfect and perfect monitoring.
In the case of imperfect monitoring, intervention schemes
can improve the suboptimal performance of non-cooperative equilibrium when the
intervention device has a sufficiently accurate monitoring technology, although
it may not be possible to achieve the best feasible performance.
In the case of perfect monitoring, the best feasible performance can
be obtained with an intervention scheme when the intervention device has a sufficiently strong intervention capability.
\end{abstract}




\section{Introduction}

When self-interested users share resources non-cooperatively, it is
common that the resources are utilized suboptimally from a global
point of view \cite{Hardin}. Hence, overcoming the
suboptimal performance of non-cooperative outcomes poses an
important challenge for successful resource utilization. The aforementioned
phenomenon is widely observed in wireless communications, where
users compete for radio resources interfering with each other. For
the sake of discussion, consider the following abstract scenario of
resource sharing in communications. First, users determine
their resource usage levels, which in turn determine the service
quality they receive. In general, as the overall usage level
increases, the service quality is reduced due to interference or
congestion. The payoff of a user is determined by its own usage
level as well as the service quality. In such a scenario, users tend
to choose a higher usage level than the socially optimal one. That
is, it is in the self-interest of users to choose a high usage
level, although reducing their usage levels simultaneously would
benefit all of them. In game theory, such a conflict between private
and social interests is modeled as the prisoner's dilemma game. In
the literature, it has been shown that various wireless
communication scenarios exhibit a prisoner's dilemma phenomenon,
including packet forwarding \cite{hubaux}, distributed spectrum
allocation \cite{laufer}, and medium access control (CSMA/CA
\cite{Cagalj} and slotted Aloha \cite{ma}).

Incentive schemes are needed to improve the performance
of non-cooperative outcomes.
In this paper, we propose a class of incentive schemes based on the
idea of intervention.
Implementing an intervention scheme requires an intervention device
that is able to monitor the actions of users and to affect their
resource usage. An intervention manager first chooses an intervention
rule used by the intervention device, and then users choose their
actions knowing the intervention rule chosen by the manager. After
observing a signal about the actions of users, the intervention
device chooses its action according to the intervention rule. The
manager chooses an intervention rule to maximize his payoff,
anticipating the rational behavior of users given the intervention
rule. The payoff of the manager can be considered as a
measure of the system performance, which can incorporate various
efficiency and fairness criteria. We formulate the interaction between
users and a manager as an intervention game and propose a solution concept called
intervention equilibrium. Intervention equilibrium predicts
the outcome of an intervention game in terms of an intervention rule
chosen by the manager and an operating point chosen by users.

\begin{figure}%
\centering
\includegraphics[width=1\textwidth]{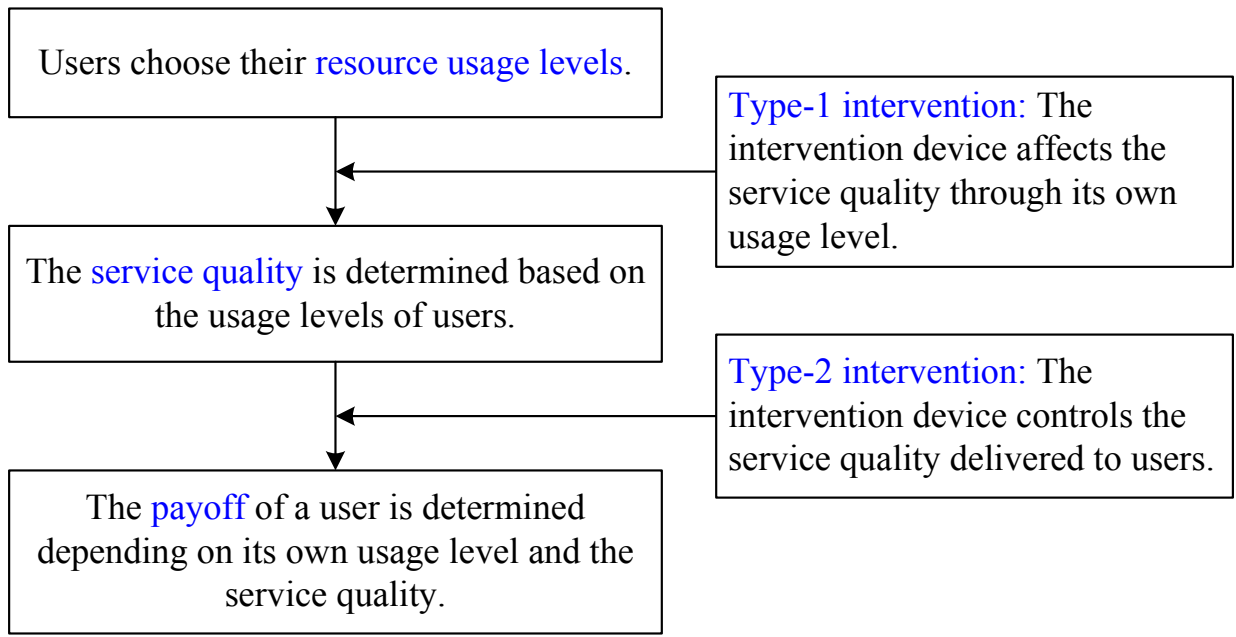}
\caption{Two types of intervention in a resource sharing scenario.}%
\label{fig:resource}
\end{figure}

Intervention can be classified into two types, called type 1 and
type 2, depending on how the intervention device acts in the system
relative to users. In type-1 intervention, the intervention device
acts in a symmetric way as users do while having the ability to
monitor the actions of other users. An example of type-1
intervention can be found in \cite{jpark} and \cite{jpark2}, which consider a random
access network where an intervention device interferes with other
users by transmitting its packets after obtaining information about the transmission
probabilities of users. In type-2 intervention, the intervention
device acts as a gatekeeper which can control resource usage by
users. An example of type-2 intervention can be found in
\cite{garg} and \cite{gai}. \cite{garg} analyzes scheduling mechanisms where a scheduler
assigns different priorities to traffic flows
depending on their input rates, and \cite{gai} considers a packet dropping
mechanism where the server determines the probability of dropping packets as a function
of the total arrival rate.
The two types of intervention can be
applied to the aforementioned resource sharing scenario, as
schematically shown in Fig.~\ref{fig:resource}.

\begin{figure}%
\centering
\includegraphics[width=1\textwidth]{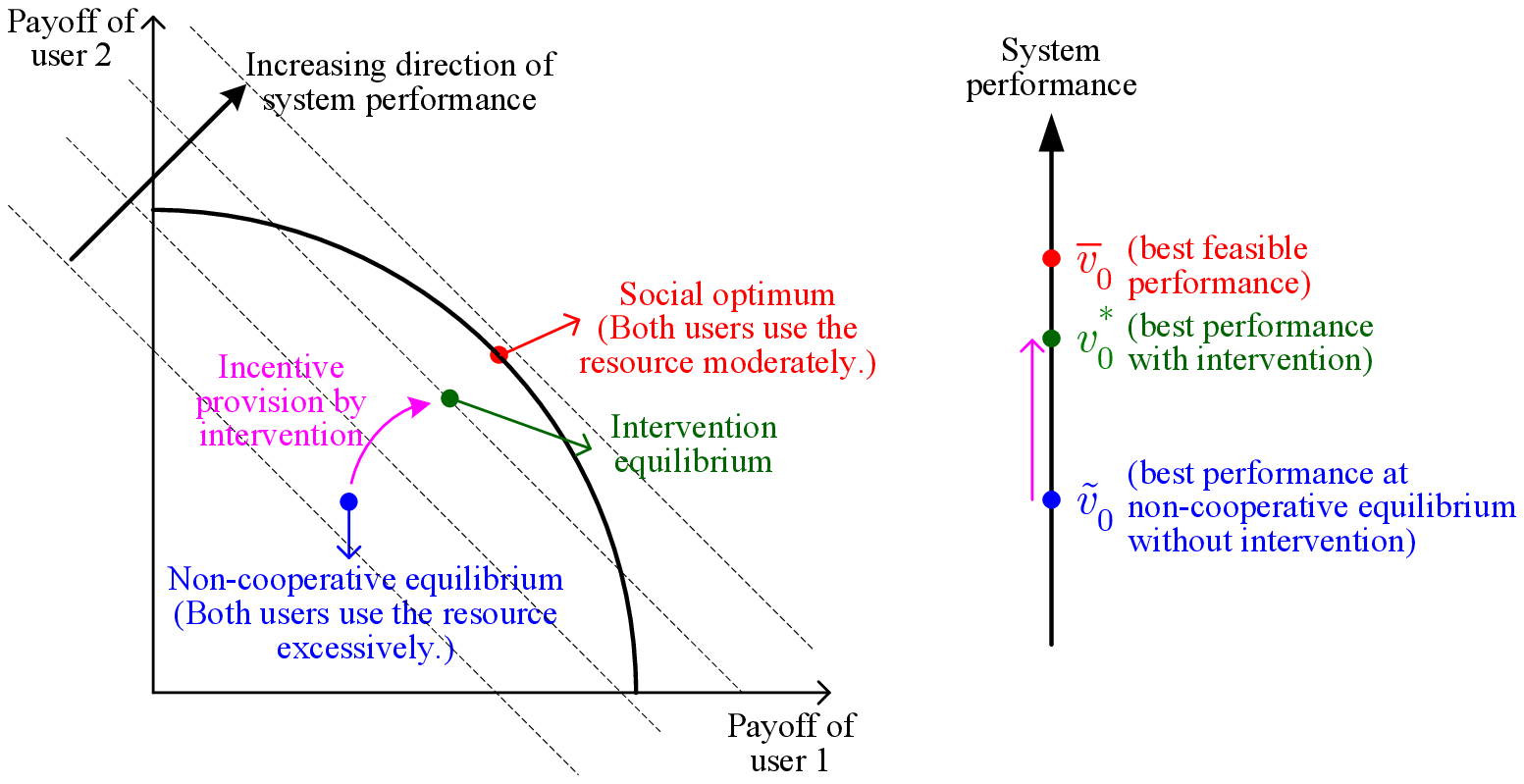}
\caption{Performance improvement through an intervention scheme. (The system performance is given
by the average payoff, and a dotted line
represents the set of payoff profiles that yield the same system performance.)}%
\label{fig:twouser}
\end{figure}

The goal of intervention schemes to improve the performance
of non-cooperative outcomes is illustrated in Fig.~\ref{fig:twouser} with
two users and the system performance measured by
the average payoff of the two users. Our analysis is aimed at answering the following
two questions.
\begin{enumerate}
\item When can we construct an intervention scheme that improves the
suboptimal performance of non-cooperative equilibrium?
\item When can we construct an intervention scheme that achieves
the best feasible performance?
\end{enumerate}
Our analysis suggests that the answers to these questions depend on the
ability of the intervention device:
\begin{itemize}
\item Ability to monitor the actions of users (i.e., monitoring technology),
\item Ability to affect the payoffs of users through its actions (i.e., intervention capability).
\end{itemize}
The discussion on the example in Section~\ref{sec:ex1} shows that
an intervention scheme can improve the performance of non-cooperative
equilibrium when the monitoring technology is sufficiently accurate.
This result is reinforced by the analytical results and the example in
Section~\ref{sec:permon}, which considers the case of perfect monitoring.
The analytical result in Section~\ref{sec:impmon} shows that intervention
schemes may not achieve the best feasible performance when the monitoring
technology is noisy. On the other hand, the analytical results and the
example in Section~\ref{sec:permon} show that intervention
schemes can achieve the best feasible performance when monitoring is perfect
and the intervention device has a sufficiently strong intervention capability.
When signals are noisy, the manager can provide incentives by
triggering a punishment following signals that are more likely to
occur when users deviate. When these signals occur with positive
probability even when users do not deviate, punishment happens from time
to time at equilibrium, which results in a performance
loss. On the contrary, when signals are perfectly accurate,
punishment through intervention can be used only as a threat, which
is never used at equilibrium. Thus, in the case of perfect
monitoring, it is possible for the manager to achieve a desired operating point without
incurring a performance loss.

The rest of this paper is organized as follows. In Section~\ref{sec:formulation},
we formulate intervention games, develop a solution concept of intervention
equilibrium, and show its existence in a finite intervention game.
In Sections~\ref{sec:noisy} and \ref{sec:permon}, we derive analytical results
and discuss illustrative examples in the cases
of imperfect and perfect monitoring, respectively.
In Section~\ref{sec:compare}, we compare intervention schemes with existing approaches in the literature.
In Section~\ref{sec:conclusion}, we conclude.

\section{Intervention Games and Intervention Equilibrium} \label{sec:formulation}

We consider a system (e.g., a wireless network) where $N$ users and an intervention device
interact. The set of the users is finite and denoted by $\mathcal{N} =
\{1,\ldots,N\}$. The action space of user $i$ is denoted by $A_i$,
and a pure action for user $i$ is denoted by $a_i \in A_i$, for all
$i \in \mathcal{N}$. A pure action profile is represented by a
vector $a = (a_1, \ldots, a_N)$, and the set of pure action
profiles is denoted by $A \triangleq \prod_{i
\in \mathcal{N}} A_i$. A mixed action for user $i$ is a probability
distribution over $A_i$ and is denoted by $\alpha_i \in
\Delta(A_i)$, where $\Delta(X)$ is the set of all probability
distributions over a set $X$. A mixed action profile is represented by a
vector $\mathbf{\alpha} = (\alpha_1, \ldots, \alpha_N) \in \prod_{i
\in \mathcal{N}} \Delta(A_i)$.
A mixed action profile of the users other than user~$i$ is written as $\alpha_{-i} = (\alpha_1,\ldots,
\alpha_{i-1},\alpha_{i+1}, \ldots,\alpha_N)$ so that $\alpha$ can be expressed as $\alpha = (\alpha_i,\alpha_{-i})$.
Once a pure action profile of the
users is determined, a signal is realized from the set of all
possible signals, denoted $Y$, and is observed by the intervention device.
We represent the probability distribution of signals by a mapping $\rho:A \rightarrow \Delta(Y)$.
That is, $\rho(a) \in \Delta(Y)$ denotes the probability distribution of signals
given a pure action profile $a$.
When $Y$ is finite, the probability that a signal $y$ is realized given a pure
action profile $a$ is denoted by $\rho(y|a)$.
After observing
the realized signal, the intervention device takes its action, called an intervention
action.
We use $a_0$, $\alpha_0$, and $A_0$ to denote a pure action, a mixed action, and
the set of pure actions for the intervention device, respectively.

Since the intervention device chooses its action after observing the signal, a
strategy for it can be represented by a mapping $f:Y \rightarrow
\Delta(A_0)$, which is called an \emph{intervention rule}.
That is, $f(y) \in \Delta(A_0)$ denotes the mixed action for the intervention
device when it observes a signal $y$.
When $A_0$ is finite, the probability that the intervention
device takes an action $a_0$ given a signal $y$ is denoted by $f(a_0|y)$.
The set of all possible intervention rules is denoted by $\mathcal{F}$.
There is a system manager who determines the intervention rule used by the intervention
device. We assume that the manager can commit to an intervention rule,
for example, by using a protocol embedded in the intervention device.
The payoffs of the users and the manager are determined by the actions of the intervention device and
the users and the realized signal. We denote the
payoff function of user $i \in \mathcal{N}$ by $u_i: A_0 \times A \times Y \rightarrow \mathbb{R}$
and that of the manager by $u_0: A_0 \times A \times Y \rightarrow \mathbb{R}$.
We call the pair $(Y,\rho)$ the \emph{monitoring technology} of the intervention device,
and call $A_0$ its \emph{intervention capability}. An intervention device is
characterized by these two, and we represent an \emph{intervention scheme} by $\langle (Y,\rho),A_0,f \rangle$.

The game played by the manager and the users is formulated as an \emph{intervention game}, which is summarized by the data
\begin{align*}
\Gamma = \left\langle \mathcal{N}_0, (A_i)_{i \in \mathcal{N}_0}, (u_i)_{i \in \mathcal{N}_0},
(Y,\rho) \right\rangle,
\end{align*}
where $\mathcal{N}_0 \triangleq \mathcal{N} \cup \{0\}$.
The sequence of events in an intervention game can be listed as follows.
\begin{enumerate}
\item The manager chooses an intervention rule $f \in \mathcal{F}$.
\item The users choose their actions $\mathbf{\alpha} \in \prod_{i \in \mathcal{N}} \Delta(A_i)$
simultaneously, knowing the intervention rule $f$ chosen by the manager.
\item A pure action profile $a$ is realized following the probability distribution $\alpha$,
and a signal $y \in Y$ is realized following the probability distribution $\rho(a)$.
\item The intervention device chooses its action $a_0 \in A_0$ following the probability
distribution $f(y)$.
\end{enumerate}

Ex ante payoffs, or expected payoffs given an intervention rule
and a pure action profile, can be computed by taking expectations with respect to signals and intervention actions.
The ex ante payoff function of user $i$ is denoted by
a function $v_i:\mathcal{F} \times A \rightarrow \mathbb{R}$,
while that of the manager is denoted by $v_0:\mathcal{F} \times A \rightarrow \mathbb{R}$.
We say that an intervention game is \emph{finite} if $A_i$, for $i \in
\mathcal{N}_0$, and $Y$ are all finite. In a finite intervention game, ex ante
payoffs can be computed as
\begin{align*}
v_i(f,a) = \sum_{y \in Y} \sum_{a_0 \in A_0} u_i(a_0,a,y) f(a_0|y) \rho(y|a),
\end{align*}
for all $i \in \mathcal{N}_0$.
Once the manager chooses an intervention rule $f$, the users play a simultaneous game, whose
normal form representation is given by
\begin{align*}
\Gamma_f = \left\langle \mathcal{N}, (A_i)_{i \in \mathcal{N}}, (v_i(f,\cdot))_{i \in \mathcal{N}}
\right\rangle.
\end{align*}
We predict actions chosen by the users given an intervention rule $f$ by applying the solution concept of Nash
equilibrium \cite{fudenberg} to the induced game $\Gamma_f$. With an abuse of notation, we extend
the domain of $v_i$ to $\mathcal{F} \times \prod_{i \in \mathcal{N}} \Delta(A_i)$ for all $i \in \mathcal{N}_0$
by taking expectation with respect to pure action profiles.

\begin{definition}
An intervention rule $f \in \mathcal{F}$ \emph{sustains} an action profile
$\alpha^* \in \prod_{i \in \mathcal{N}} \Delta(A_i)$ if $\alpha^*$ is a Nash equilibrium of
the game $\Gamma_f$, i.e.,
\begin{align*}
v_i(f,\alpha_i^*, \alpha_{-i}^*) \geq v_i(f,\alpha_i,\alpha_{-i}^*) \quad \text{for all $\alpha_i \in
\Delta(A_i)$, for all $i \in \mathcal{N}$}.
\end{align*}
An action profile $\alpha^*$ is \emph{sustainable} if there exists
an intervention rule $f$ that sustains $\alpha^*$.
\end{definition}

Let $\mathcal{E}(f) \subseteq \prod_{i \in \mathcal{N}} \Delta(A_i)$ be the set of action profiles
sustained by $f$. We say that a pair $(f, \alpha)$ is attainable if $\alpha \in \mathcal{E}(f)$.
The manager's problem is to find an attainable pair that maximizes his ex ante payoff among all
attainable pairs, which leads to the following solution concept for intervention games.
\begin{definition} \label{def:ie}
$(f^*,\alpha^*) \in \mathcal{F} \times \prod_{i \in \mathcal{N}} \Delta(A_i)$ is an \emph{intervention equilibrium}
if $\alpha^* \in \mathcal{E}(f^*)$ and
\begin{align*}
v_0(f^*, \alpha^*) \geq v_0(f, \alpha) \quad \text{for all $(f,\alpha)$ such that $\alpha \in \mathcal{E}(f)$.}
\end{align*}
$f^* \in \mathcal{F}$ is an \emph{optimal intervention rule} if there exists
an action profile $\alpha^* \in \prod_{i \in \mathcal{N}} \Delta(A_i)$ such that $(f^*,\alpha^*)$ is an
intervention equilibrium.
\end{definition}

An intervention equilibrium solves the following optimization problem:
\begin{align} \label{eq:optint}
\max_{(f,\alpha)} v_0(f,\alpha) \text{ subject to } \alpha \in \mathcal{E}(f).
\end{align}
The constraint $\alpha \in \mathcal{E}(f)$ represents incentive constraints for the users,
which require that the users choose the action profile $\alpha$ in their self-interest
given the intervention rule $f$. The problem \eqref{eq:optint} can be rewritten as
$\max_{f \in \mathcal{F}} \max_{\alpha \in \mathcal{E}(f)} v_0(f,\alpha)$.
Then an intervention equilibrium can be considered as a subgame
perfect equilibrium (or Stackelberg equilibrium), with an implicit assumption that the manager
can induce the users to choose the best Nash equilibrium for him in case of multiple Nash
equilibria. Our interpretation is that, in order to achieve an intervention
equilibrium $(f^*,\alpha^*)$, the manager announces the intervention rule $f^*$
and recommends the action profile $\alpha^*$ to the users. Since $\alpha^* \in \mathcal{E}(f^*)$,
the users do not have an incentive to deviate unilaterally from $\alpha^*$, and
$\alpha^*$ becomes a focal point \cite{fudenberg} of the game $\Gamma_{f^*}$.
Below we show the existence of an intervention equilibrium in a finite intervention game.

\begin{proposition} \label{prop:existence}
Every finite intervention game has an intervention equilibrium.
\end{proposition}

We prove Proposition~\ref{prop:existence} using the following two lemmas.

\begin{lemma} \label{lemma:euhc}
The correspondence $\mathcal{E}: \mathcal{F} \rightrightarrows \prod_{i \in \mathcal{N}} \Delta(A_i)$ is
nonempty, compact-valued, and upper hemi-continuous.
\end{lemma}

\begin{IEEEproof}
We can show that, for any $f \in \mathcal{F}$, the set $\mathcal{E}(f)$ is nonempty by
applying Nash Theorem \cite{nash} to $\Gamma_f$. Since $\prod_{i \in \mathcal{N}} \Delta(A_i)$
is bounded, it suffices to show that $\mathcal{E}$ has a closed graph
to prove that $\mathcal{E}$ is compact-valued and upper hemi-continuous (u.h.c.) (see Theorem 3.4 of \cite{stokey}).
Choose a sequence $\{(f^n,\alpha^n)\}$ with $(f^n,\alpha^n) \rightarrow (f,\alpha)$
and $\alpha^n \in \mathcal{E}(f^n)$ for all $n$. Suppose that $\alpha \notin \mathcal{E}(f)$.
Then there exists $i \in \mathcal{N}$ such that $\alpha_i$ is not a best response
to $\alpha_{-i}$ in $\Gamma_f$. Then there exist $\epsilon > 0$ and $\alpha'_i$ such
that $v_i(f,\alpha'_i,\alpha_{-i}) > v_i(f,\alpha_i,\alpha_{-i}) + 3\epsilon$. Since $v_i$
is continuous and $(f^n,\alpha^n) \rightarrow (f,\alpha)$, for sufficiently large $n$ we have
\begin{align*}
v_i(f^n,\alpha'_i,\alpha_{-i}^n) > v_i(f,\alpha'_i,\alpha_{-i}) - \epsilon
> v_i(f,\alpha_i,\alpha_{-i}) + 2\epsilon > v_i(f^n,\alpha_i^n,\alpha_{-i}^n) + \epsilon,
\end{align*}
which contradicts $\alpha^n \in \mathcal{E}(f^n)$.
\end{IEEEproof}

Define a function $\check{v}_0:\mathcal{F} \rightarrow \mathbb{R}$ by
$\check{v}_0(f) = \max_{\alpha \in \mathcal{E}(f)} v_0(f,\alpha)$.
For each $f$, $\mathcal{E}(f)$ is nonempty and compact by Lemma~\ref{lemma:euhc} and
$v_0(f,\cdot)$ is continuous. Hence, the function $\check{v}_0$ is
well-defined.

\begin{lemma} \label{lemma:v0usc}
The function $\check{v}_0$ is
upper semi-continuous.
\end{lemma}

\begin{IEEEproof}
Let $E(f) = \{ \alpha \in \mathcal{E}(f) : v_0(f,\alpha) = \check{v}_0(f) \}$.
Note that $E(f)$ is nonempty for all $f$.
Fix $f$, and let $\{f^n\}$ be any sequence converging to $f$.
Choose $\alpha^n \in E(f^n)$, for all $n$. Let $v_0^s = \limsup_{n \rightarrow \infty}
\check{v}_0(f^n)$. Then there exists a subsequence $\{f^{n_k}\}$ such that
$v_0^s = \lim v_0(f^{n_k},\alpha^{n_k})$. Since $\alpha^n \in \mathcal{E}(f^n)$
and $\mathcal{E}$ is u.h.c., there exists a convergent subsequence of $\{\alpha^{n_k}\}$,
called $\{\alpha^j\}$, whose limit point $\alpha$ is in $\mathcal{E}(f)$.
Hence, $v_0^s = \lim v_0(f^{j},\alpha^{j}) = v_0(f,\alpha) \leq \check{v}_0(f)$
since $\alpha \in \mathcal{E}(f)$.
\end{IEEEproof}

Note that the space of intervention rules, $\mathcal{F}$,
is equivalent to $(\Delta(A_0))^{|Y|}$, which is compact.
Therefore, a solution to $\max_{f \in \mathcal{F}} \check{v}_0(f)$ exists,
which establishes the existence of an intervention equilibrium.
This completes the proof of Proposition~\ref{prop:existence}.

There can be multiple intervention equilibria, all of which yield the same
payoff for the manager. We can propose different selection criteria for
the manager to choose an intervention equilibrium out of multiple ones. For example,
the discussion on affine intervention rules in Section~\ref{sec:permonanal} is
motivated by the robustness of performance to mistakes by the users as well as
simplicity.

Recall that an intervention device is characterized by $(Y,\rho)$ and $A_0$.
In this paper, we focus on the problem of finding an optimal intervention rule
when the manager has a particular intervention device. However, we can think of
a scenario where the manager can select an intervention device from multiple ones
given the operating cost of each available intervention device.
Our analysis in this paper allows the manager to evaluate the optimal performance
achieved with each intervention device. He can then select the best intervention
device taking into account both performance and cost.

\section{Performance with Intervention under Imperfect Monitoring} \label{sec:noisy}

\subsection{Analytical Results} \label{sec:impmon}

In this section, we maintain the following assumption.
\begin{assumption} \label{ass:tilde}
There exists an action for the intervention device $\tilde{a}_0 \in A_0$ that satisfies
\begin{align*}
u_0(\tilde{a}_0,a,y) > u_0(a_0,a,y) \quad \text{for all $a_0 \neq \tilde{a}_0$},
\end{align*}
for all $a \in A$ and $y \in Y$.
\end{assumption}

Assumption~\ref{ass:tilde} asserts the existence of an intervention action
that is most preferred by the manager regardless of the action profile of the users and the signal.
We can interpret the most preferred intervention action, $\tilde{a}_0$, as the
intervention action that corresponds to no intervention. Then Assumption~\ref{ass:tilde}
states that exerting intervention is costly for the manager, reflecting that
intervention typically degrades the overall performance. Moreover, there is some
operational cost (e.g., energy consumption) needed to exert intervention.

Define an intervention rule $\tilde{f}$ by $\tilde{f}(y) = \tilde{a}_0$ for all $y$.
It can be considered that the manager decides not to intervene at all when he chooses $\tilde{f}$.
Let $\overline{v}_0 = \sup_{(f,\alpha)} v_0(f,\alpha)$, $v_0^* = \sup_{f}
\sup_{\alpha \in \mathcal{E}(f)} v_0(f,\alpha)$, and $\tilde{v}_0 = \sup_{\alpha \in
\mathcal{E}(\tilde{f})} v_0(\tilde{f},\alpha)$. $\overline{v}_0$ is the best performance that the manager
can obtain when the users are not subject to the incentive constraints (e.g., when the
actions of the users can be completely controlled by the manager).
$v_0^*$ is the best performance when the manager is required to satisfy
the incentive constraints for the users.
Lastly, $\tilde{v}_0$ is the best performance when the manager does not
engage in active intervention. It is straightforward to see that $\tilde{v}_0 \leq v_0^* \leq \overline{v}_0$.
The following proposition provides a sufficient condition
on the intervention game for a gap between $\overline{v}_0$ and $v_0^*$ to exist.

\begin{proposition} \label{thm:ineff}
Suppose that the intervention game is finite, $\rho$ has full support (i.e., $\rho(y|a) > 0$
for all $y$ and $a$), and there is no $\alpha$ such that $\alpha \in \mathcal{E}(\tilde{f})$
and $v_0(\tilde{f},\alpha) = \overline{v}_0$. Then $v_0^* < \overline{v}_0$.
\end{proposition}

\begin{IEEEproof}
Suppose that the conclusion does not hold, i.e., $v_0^* = \overline{v}_0$.
Since the intervention game is finite, $v_0^*$ is attained by Proposition~\ref{lemma:euhc}. Thus, there exists
$(f,\alpha)$ such that $\alpha \in \mathcal{E}(f)$ and $v_0(f,\alpha) =
\overline{v}_0$. Note that $\overline{v}_0 = v_0(f,\alpha) \leq v_0(\tilde{f},\alpha) \leq \overline{v}_0$.
Hence, $v_0(f,\alpha) = v_0(\tilde{f},\alpha)$. Since $\rho$ has full support,
we have $f(y) = \tilde{f}(y)$ for all $y$. This contradicts the hypothesis
that there is no $\alpha$ such that $\alpha \in \mathcal{E}(\tilde{f})$
and $v_0(\tilde{f},\alpha) = \overline{v}_0$.
\end{IEEEproof}

When the intervention game is finite, $\overline{v}_0$ is attained since $v_0$
is continuous and $(\mathcal{F} \times \prod_{i \in \mathcal{N}} \Delta(A_i))$
is compact. Since $v_0(\tilde{f},\alpha) \geq v_0(f,\alpha)$ for all $\alpha$,
for all $f$, we have $\overline{v}_0 = \max_{\alpha} v_0(\tilde{f},\alpha)$.
In fact, when the intervention game is finite and $\rho$ has full support, $\tilde{f}$ is the only intervention rule
that can attain the best feasible performance, $\overline{v}_0$. When $\tilde{f}$ sustains
no action profile that attains $\overline{v}_0$, the manager needs to trigger a punishment
following some signals in order to provide appropriate incentives for the users
to follow an action profile such that $v_0(\tilde{f},\alpha) = \overline{v}_0$. However, since $\rho$ has full support,
the punishment results in a performance loss, which prevents the manager
from achieving $\overline{v}_0$.

\subsection{Illustrative Example (Type-2 Intervention)} \label{sec:ex1}

\begin{figure}%
\centering
\includegraphics[width=0.9\textwidth]{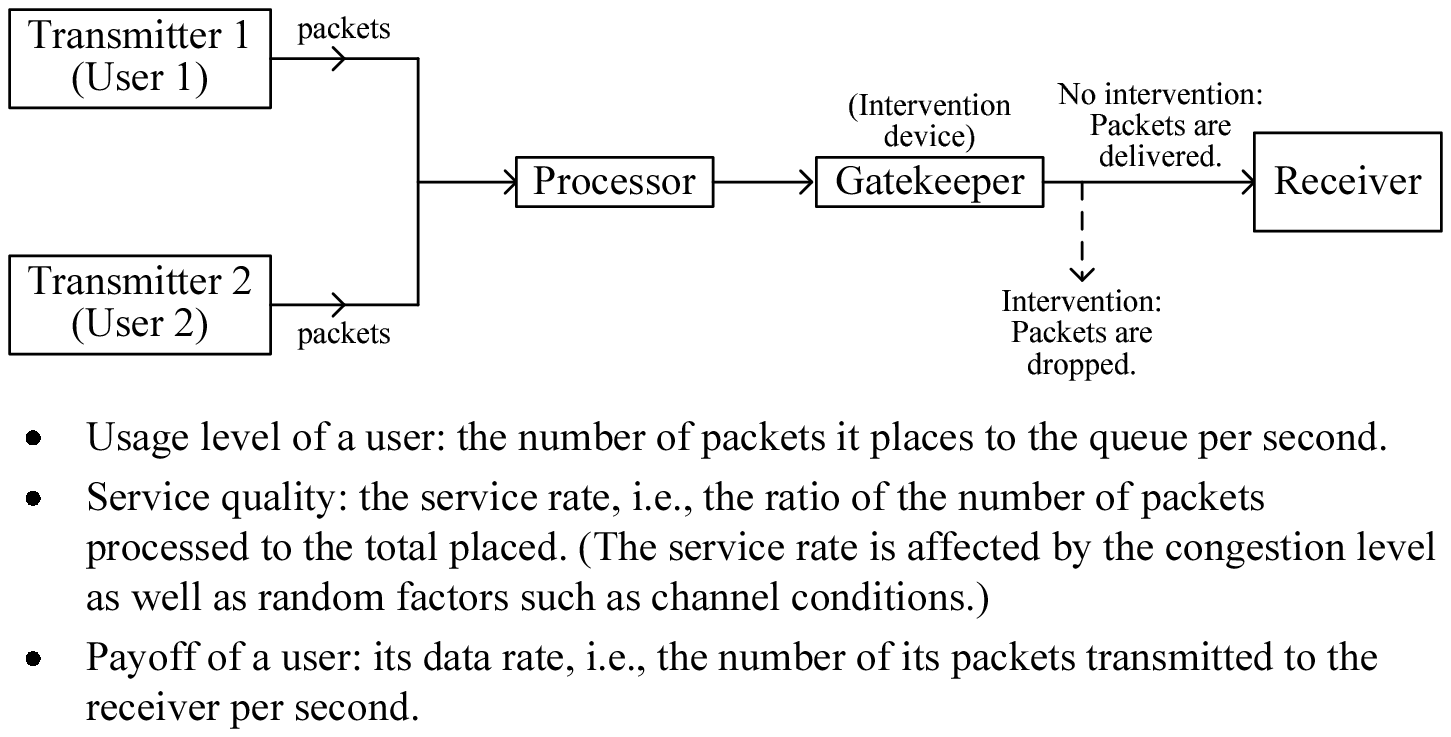}
\caption{A communication scenario that fits into the example in
Section~\ref{sec:ex1}.} \label{fig:type2}
\end{figure}

We consider a wireless network where two users interfere with each other.
Each user has two pure actions, $a_L$ and $a_H$, which represent low and high resource usage levels, respectively,
and satisfy $0 < a_L < a_H$.
The service quality is determined randomly given an action profile, and there are
two possible quality levels, $\overline{y}$ and $\underline{y}$, with $0 < \underline{y} < \overline{y}$.
The service quality is realized following the distribution
\begin{align*}
\rho(\overline{y}|a) = \left\{
\begin{array}{ll}
p, \quad &\text{if $a=(a_L,a_L)$,}\\
q, \quad &\text{if $a=(a_H,a_L)$ or $(a_L,a_H)$,}\\
r, \quad &\text{if $a=(a_H,a_H)$,}
\end{array} \right.
\end{align*}
where $0 < r < q < p < 1$.
The intervention device in this example acts as a gatekeeper (i.e., type-2 intervention) after observing the service
quality, having two
pure actions: intervene ($\hat{a}_0$) and not intervene ($\tilde{a}_0$).
When the intervention
device does not intervene, a user receives a payoff given by the product of the quality level and
its own usage level, i.e., $u_i(\tilde{a}_0,a,y) = y a_i$ for all $a$ and $y$, for $i=1,2$.
When the intervention device does intervene, the service stops completely and
a user receives zero payoff regardless of its usage level, i.e., $u_i(\hat{a}_0,a,y) = 0$ for
all $a$ and $y$, for $i=1,2$.
The payoff of the manager is set as the average payoff of the users,
i.e., $u_0(a_0,a,y) = [u_1(a_0,a,y) + u_2(a_0,a,y)]/2$.
Note that denoting the action of not intervening by
$\tilde{a}_0$ is consistent with Assumption~\ref{ass:tilde}.
A communication scenario that fits into this example is presented in Fig.~\ref{fig:type2}.

Since there are only two pure actions for the intervention device, we can represent $\mathcal{F} = [0,1]$ and use $f(y)$ as the probability
of not intervening given the signal $y$.
The ex ante payoff function of user $i$ is given by
\begin{align*}
v_i(f,a) = [\rho(\overline{y}|a)f(\overline{y})\overline{y} + (1-\rho(\overline{y}|a))f(\underline{y})\underline{y}]a_i.
\end{align*}
The payoff matrix of the game $\Gamma_{\tilde{f}}$, i.e., the game when the intervention
device does not intervene at all, is displayed in Table~\ref{table:pd}, where we
define $y_k = k \overline{y} + (1-k) \underline{y}$, for $k=p,q,r$. We assume that the
game $\Gamma_{\tilde{f}}$ is the prisoner's dilemma game, i.e., $y_q a_H > y_p a_L
> y_r a_H > y_q a_L$ and $2 y_p a_L > y_q (a_H + a_L)$. Then without any intervention,
it is the dominant strategy of each user to choose the high usage level, which
results in the inefficient Nash equilibrium. The manager aims to improve
the inefficiency of the Nash equilibrium by providing appropriate incentives through
intervention.\footnote{In this paper, we focus on the role of intervention schemes
to improve the prospect of cooperation by applying intervention to prisoner's dilemma
situations. Intervention schemes can also be used to help users achieve coordination
by eliminating the multiplicity of Nash equilibria in coordination games such as
the battle of the sexes and the stag hunt \cite{fudenberg}. For example, in the stag-hunt
game, an intervention scheme may induce players to choose the payoff dominant (but
not risk dominant) ``all stag'' equilibrium by intervening in the hare hunt.}
We restrict attention to symmetric action profile, assuming that
the manager desires to sustain a symmetric action profile.

Let $w_0(\alpha) = \sup_f \{v_0(f,\alpha) : \alpha \in
\mathcal{E}(f) \}$. That is, $w_0(\alpha)$ is the maximum payoff
that the manager can obtain while sustaining a given action profile
$\alpha$. Since we focus on symmetric action profiles and there are
only two pure actions for each user, let $\alpha \in [0,1]$ denote the
probability of each user playing $a_L$. Then we can show that
$w_0(0) = y_r a_H$ and, for
$\alpha \in (0,1]$,
\begin{align} \label{eq:w0alpha}
w_0(\alpha) = \left\{
\begin{array}{l}
\frac{\{(q-r) + \alpha [(p-q) - (q-r)]\}a_H a_L}{[(1-r) a_H - (1-q)
a_L] + \alpha [(p a_L - q a_H) - (q a_L - r a_H)]}
\overline{y},\\ \qquad \qquad \qquad \text{if $\alpha (p a_L - q a_H) + (1-\alpha) (q a_L - r a_H) \geq 0$,}\\
0, \quad \qquad \qquad \text{otherwise.}
\end{array} \right.
\end{align}
The intervention rule that attains $w_0(0)$ is given by $\tilde{f}$ (i.e.,
no intervention), while the intervention rule that attains $w_0(\alpha)$, for
$\alpha \in (0,1]$, is given by
\begin{align*}
f(\overline{y}) = 1 \quad \text{and} \quad f(\underline{y}) = \frac{(q a_L - r a_H) + \alpha [(p a_L - q a_H) -
(q a_L - r a_H)]}{[(1-r) a_H - (1-q) a_L] + \alpha [(p a_L - q a_H) - (q a_L - r a_H)]} \frac{\overline{y}}{\underline{y}}
\end{align*}
if $\alpha (p a_L - q a_H) + (1-\alpha) (q a_L - r a_H) \geq 0$, and by $f(\overline{y})=f(\underline{y})=0$
otherwise. We can think of $\alpha (p a_L - q a_H) + (1-\alpha) (q a_L - r a_H)$ as
a measure of the sensitivity of signals between the two pure actions when the other
user plays $\alpha$. When signals are sufficiently sensitive at $\alpha$, an intervention rule
can sustain $\alpha$ with a positive payoff by degrading the low quality only. On the contrary,
when signals are not sensitive, destroying all the payoffs is the only
method to sustain $\alpha$, which yields zero payoff for the users.
Note that, when $p a_L - q a_H < 0$ and $q a_L - r a_H > 0$, the pure action
profile $(a_L,a_L)$ cannot be sustained with a positive payoff while a completely mixed action profile
can be. In this case, signals are more sensitive to the action of a user
when the other user plays $a_H$. Hence, by inducing the users to play $a_H$ with
positive probability, the manager can make the signal a more informative indicator
of a deviation. This allows the possibility that an intervention rule
improves the performance of non-cooperative equilibrium by sustaining
a completely mixed action profile even when the social optimum $(a_L,a_L)$ cannot be sustained non-trivially.
A similar discussion about the advantage of using mixed actions can be found
in \cite{kandori} in the context of the repeated prisoner's dilemma game.

In this example, we have $\tilde{v}_0 = w_0(0) = y_r a_H$, $\overline{v}_0 = y_p a_L$,
and $v_0^* = \max_{\alpha \in [0,1]} w_0(\alpha)$. We summarize the results
about the performance with intervention, $v_0^*$, in the following proposition.

\begin{proposition} \label{prop:ex1}
(i) Suppose that (a) $p a_L - q a_H < 0$ and $q a_L - r a_H < 0$, or (b)
$p a_L - q a_H < q a_L - r a_H$ and $(p-q)(1-r) - (q-r)(1-q) \leq 0$.
Then $v_0^* = \tilde{v}_0$.\\
(ii) Suppose that (c) $p a_L - q a_H \geq q a_L - r a_H \geq 0$, (d)
$p a_L - q a_H \geq 0 > q a_L - r a_H$, or (e) $0 \leq p a_L - q a_H < q a_L - r a_H$
and $(p-q)(1-r) - (q-r)(1-q) > 0$. Then $v_0^* = \max \{\tilde{v}_0, w_0(1) \}$.\\
(iii) Suppose that (f) $p a_L - q a_H < 0 \leq q a_L - r a_H$
and $(p-q)(1-r) - (q-r)(1-q) > 0$. Then $v_0^* = \max \{\tilde{v}_0, w_0(\overline{\alpha}) \}$, where
\begin{align*}
\overline{\alpha} = \frac{q a_L - r a_H}{(q a_L - r a_H) - (p a_L - q a_H)}.
\end{align*}
\end{proposition}

\begin{IEEEproof}
See Appendix~\ref{app:ex1}.
\end{IEEEproof}

Fig.~\ref{fig:w0} shows that each of the three cases of $v_0^* = \tilde{v}_0$, $v_0^* = w_0(\overline{\alpha})$, and
$v_0^* = w_0(1)$ can arise depending on the parameter values. To
obtain the results, we set $a_L = 1$, $a_H = 1.19$, $\overline{y} =
5$, $\underline{y} = 1$, $q = 0.8$, and $r = 0.65$ while varying
$p = 0.9, 0.94, 0.96$. We can see that, as $p$ increases, the performance
with intervention improves, getting closer to its upper bound $\overline{v}_0$.
In fact, when $v_0^* = w_0(1)$, we have
\begin{align*}
\overline{v}_0 - v_0^* = \frac{(1-p)a_L (y_q a_H - y_p a_L)}{(1-q) a_H - (1-p) a_L} > 0,
\end{align*}
which is consistent with Proposition~\ref{thm:ineff}. The gap between $v_0^*$ and $\overline{v}_0$
vanishes as $p$ approaches 1, while it increases with the deviation gain
($y_q a_H - y_p a_L$). This result is intuitive because punishment rarely occurs when $p$ is close
to 1 while a stronger punishment is needed as the deviation gain is larger.

We can consider pricing schemes applied to this example, by having the manager
charge different payments depending on the realized service quality. In order
to find a pricing scheme that sustains a certain action profile, the manager
needs to know how payments affect the payoffs of the users (i.e., the function
$u_i(a_0,a,y)$, where $a_0$ is now interpreted as the charged payments).
Suppose, for example, that the payoff of each user is given by its data rates.
Since intervention influences data rates directly, it is relatively easy
to find out how intervention actions affect payoffs.
In contrast, finding out how payments affect payoffs requires the manager to know
how the users value payments relative to data rates.
This information is difficult to obtain since the users' valuations
are subjective and thus not easily measurable. This discussion points out
the informational advantage of intervention over pricing.

\section{Performance with Intervention under Perfect Monitoring} \label{sec:permon}

\subsection{Analytical Results} \label{sec:permonanal}

In this section, we consider the case where the intervention device
can observe the pure action profile without errors (i.e., perfect monitoring),
as stated formally in the following assumption.

\begin{assumption} \label{ass:permon}
$Y = A$, and only signal $a$ can arise in the distribution $\rho(a)$ for all $a \in A$.
\end{assumption}

With Assumption~\ref{ass:permon}, we always have $y = a$, and thus we write
the payoff functions more compactly as $u_i(a_0,a)$ instead of $u_i(a_0,a,a)$, for all $i \in \mathcal{N}_0$.
We also maintain the following two assumptions in this section.

\begin{assumption} \label{ass:aligned}
There exists an action for the intervention device $\underline{a}_0 \in A_0$ that satisfies,
for all $i \in \mathcal{N}_0$,
\begin{align} \label{eq:a0order}
u_i(\underline{a}_0,a) \geq u_i(a_0,a) \quad \text{for all $a_0 \in A_0$, for all $a \in A$}.
\end{align}
\end{assumption}

\begin{assumption} \label{ass:compact}
$A_0$ is compact, and $u_i:A_0 \times A \rightarrow \mathbb{R}$ is continuous for all $i \in \mathcal{N}_0$.
\end{assumption}

Assumption~\ref{ass:aligned} states that there exists an intervention action
that is most preferred by the users and the manager regardless of the action
profile of the users. We can interpret $\underline{a}_0$ in Assumption~\ref{ass:aligned}
as the intervention action corresponding to
no intervention, similarly to $\tilde{a}_0$ in Assumption~\ref{ass:tilde}.
Then Assumption~\ref{ass:aligned} implies that intervention can only reduce the payoffs of
the users and the manager.

In this section, we restrict attention to pure actions (both for the users and for the intervention device)
while allowing the action spaces to be continuous spaces. Thus, an intervention rule
is represented by a mapping $f:A \rightarrow A_0$, while user $i$ chooses a pure
action $a_i \in A_i$ given an intervention rule.
Then the ex ante payoff function is given by $v_i(f,a) = u_i(f(a),a)$,
for all $i \in \mathcal{N}_0$. We define a class of intervention rules.

\begin{definition}
$f_{\tilde{a}}:A \rightarrow A_0$ is an \emph{extreme intervention
rule with target action profile $\tilde{a} \in A$} if $f_{\tilde{a}}$
satisfies
\begin{itemize}
\item $f_{\tilde{a}}(a) \in \arg \min_{a_0 \in A_0} u_i(a_0,a)$ if $\exists \ i \in \mathcal{N}$ such that
$a_i \neq \tilde{a}_i$ and $a_j = \tilde{a}_j$ $\forall j \neq i$, and
\item $f_{\tilde{a}}(a) = \underline{a}_0$ otherwise.
\end{itemize}
\end{definition}

By Assumption~\ref{ass:compact}, $\arg \min_{a_0 \in A_0} u_i(a_0,a)$ is non-empty
for all $a \in A$ and $i \in \mathcal{N}$. Thus, for every $\tilde{a} \in A$,
there exists an extreme intervention rule with target action profile $\tilde{a}$.
An extreme intervention rule prescribes an intervention action that minimizes
the payoff of the deviator if there is a unilateral
deviation from the target action profile while prescribing no intervention
if there is no unilateral deviation. Hence, an extreme intervention rule
provides the strongest incentive for the users to follow a given target
action profile.
Let $\mathcal{E}(\mathcal{F}) = \cup_{f \in \mathcal{F}} \mathcal{E}(f)$. That is,
$\mathcal{E}(\mathcal{F})$ is the set of all sustainable action profiles.

\begin{lemma} \label{lem:ext}
If $a^* \in \mathcal{E}(\mathcal{F})$, then $a^* \in \mathcal{E}(f_{a^*})$.
\end{lemma}

\begin{IEEEproof}
Suppose that $a^* \in \mathcal{E}(\mathcal{F})$. Then there exists an
intervention rule $f$ such that $v_i(f, a^*) \geq v_i(f,
a_i, {a}_{-i}^*)$ for all $a_i \in A_i$, for all $i \in \mathcal{N}$. Then we
obtain $v_i(f_{a^*}, a^*) = u_i(\underline{a}_0,{a}^*) \geq u_i(f({a}^*), {a}^*) \geq u_i(f(a_i, {a}_{-i}^*),
a_i, {a}_{-i}^*) \geq u_i(f_{a^*}(a_i, {a}_{-i}^*), a_i, {a}_{-i}^*) = v_i(f_{a^*}, a_i, a_{-i}^*)$ for all $a_i
\neq a_i^*$, for all $i \in \mathcal{N}$, where the first inequality follows from \eqref{eq:a0order}
and the third from the definition of extreme intervention rules.
\end{IEEEproof}

Let $\mathcal{E}^* = \{ a \in A : a \in \mathcal{E}(f_a) \}$.
The following results are the consequences of Lemma~\ref{lem:ext}.

\begin{proposition} \label{thm:char}
(i) $\mathcal{E}(\mathcal{F}) = \mathcal{E}^*$.\\
(ii) If $(f^*,a^*)$ is an intervention equilibrium, then $(f_{a^*},
a^*)$ is also an intervention equilibrium.
\end{proposition}

\begin{IEEEproof}
(i) Let $a^* \in \mathcal{E}^*$. Then $a^* \in \mathcal{E}(f_{a^*}) \subset \mathcal{E}(\mathcal{F})$.
Hence, $\mathcal{E}^* \subset \mathcal{E}(\mathcal{F})$. The other inclusion
$\mathcal{E}(\mathcal{F}) \subset \mathcal{E}^*$ follows from Lemma~\ref{lem:ext}.

(ii) Suppose that $(f^*,a^*)$ is an intervention equilibrium. Then
by Definition~\ref{def:ie}, $a^* \in \mathcal{E}(f^*)$ and
$v_0(f^*,a^*) \geq v_0(f,a)$ for all $(f,a) \in \mathcal{F} \times A$ such
that $a \in \mathcal{E}(f)$. Since ${a}^* \in \mathcal{E}(\mathcal{F})$,
$a^* \in \mathcal{E}(f_{a^*})$ by Lemma~\ref{lem:ext}. Hence, $v_0(f^*,{a}^*) \geq
v_0(f_{a^*},a^*)$. On the other hand, since $f_{a^*}(a^*) =
\underline{a}_0$, we have $v_0(f^*,a^*) \leq
v_0(f_{a^*},a^*)$ by \eqref{eq:a0order}. Therefore,
$v_0(f^*,a^*) = v_0(f_{a^*},a^*)$, and thus
$v_0(f_{a^*},a^*) \geq v_0(f,a)$ for all $(f,a) \in \mathcal{F} \times A$ such
that $a \in \mathcal{E}(f)$. This proves that $(f_{a^*}, a^*)$ is
an intervention equilibrium.
\end{IEEEproof}

Proposition~\ref{thm:char} shows that it is without loss of generality
to restrict attention to pairs of the form $(f_a, a)$ when we ask whether a
given action profile is sustainable and whether there exists an intervention equilibrium.
The basic idea is that, in order to sustain an action profile,
it suffices to consider an intervention rule that punishes a deviator
most severely.
The role of extreme intervention rules is analogous to
that of optimal penal codes \cite{abreu} in repeated games with perfect
monitoring.
The following proposition characterizes intervention equilibria among
pairs of the form $(f_a, a)$.

\begin{proposition} \label{prop:charie}
$(f_{a^*}, a^*)$ is an intervention equilibrium if and only if $a^*
\in \mathcal{E}^*$ and $u_0(\underline{a}_0, a^*) \geq
u_0(\underline{a}_0, a)$ for all $a \in \mathcal{E}^*$.
\end{proposition}

\begin{IEEEproof}
Suppose that $(f_{a^*}, a^*)$ is an intervention equilibrium. Then
$a^* \in \mathcal{E}(f_{a^*})$, and thus $a^* \in \mathcal{E}^*$. Also,
$v_0(f_{a^*}, a^*) \geq v_0(f,a)$ for all $(f,a)$ such that $a \in \mathcal{E}(f)$.
Choose any $a \in \mathcal{E}^*$. Then $a \in \mathcal{E}(f_a)$, and thus $u_0(\underline{a}_0, a^*) =
v_0(f_{a^*},a^*) \geq v_0(f_a,a) = u_0(\underline{a}_0, a)$.

Suppose that $a^* \in \mathcal{E}^*$ and $u_0(\underline{a}_0, a^*) \geq
u_0(\underline{a}_0, a)$ for all $a \in \mathcal{E}^*$. To prove
that $(f_{a^*}, a^*)$ is an intervention equilibrium, we need to show that
(i) $a^* \in \mathcal{E}(f_{a^*})$, and (ii) $v_0(f_{a^*}, a^*) \geq v_0(f,a)$ for all $(f,a)$
such that $a \in \mathcal{E}(f)$. (i) follows from $a^* \in \mathcal{E}^*$. To prove
(ii), choose any $(f,a)$ such that $a \in \mathcal{E}(f)$. By Lemma~\ref{lem:ext},
we have $a \in \mathcal{E}^*$.
Then $v_0(f_{a^*},a^*) = u_0(\underline{a}_0, a^*) \geq
u_0(\underline{a}_0, a) \geq v_0(f,a)$, where the first inequality
follows from $a \in \mathcal{E}^*$.
\end{IEEEproof}

Proposition~\ref{prop:charie} shows that the pair $(f_a, a)$ constitutes an intervention equilibrium if $a$ solves
\begin{align} \label{eq:charie}
\max_{a \in \mathcal{E}^*} u_0(\underline{a}_0,a).
\end{align}
The next proposition provides a sufficient condition under which an intervention equilibrium exists.

\begin{proposition} \label{prop:exist2}
If $A_i$ is a bounded set in Euclidean space for all $i \in \mathcal{N}$,
then there exists an intervention equilibrium.
\end{proposition}

\begin{IEEEproof}
By Proposition~\ref{thm:char}(ii) and Proposition~\ref{prop:charie}, an intervention equilibrium exists
if and only if there exists a solution to the problem \eqref{eq:charie}. Since $u_0(\underline{a}_0,a)$
is continuous in $a$, the result follows if we show that the constraint set $\mathcal{E}^*$
is compact. Since $\mathcal{E}^* \subset A$ and $A$ is bounded, $\mathcal{E}^*$ is also bounded.
Let $G_i(a) \triangleq \arg \min_{a_0 \in A_0} u_i(a_0,a)$ for all $a \in A$, for all $i \in \mathcal{N}$.
By the Theorem of the Maximum \cite{stokey}, $G_i(a)$ is compact-valued and u.h.c.
To show that $\mathcal{E}^*$ is closed, choose a sequence $\{a^n\}$ with $a^n \rightarrow a^*$
and $a^n \in \mathcal{E}^*$ for all $n$.
Choose any $i \in \mathcal{N}$ and $a'_i \in A_i$.
Let $\{a_0^n\}$ be a sequence such that $a_0^n \in G_i(a'_i,a^n_{-i})$ for all $n$.
Since $a^n \in \mathcal{E}(f_{a^n})$, we have $u_i(\underline{a}_0,a^n) \geq u_i(a_0^n,a'_i,a^n_{-i})$.
Also, since $G_i(a)$ is u.h.c., there exists a convergent subsequence of $\{a_0^n\}$
whose limit point $a_0^*$ is in $G_i(a'_i,a^*_{-i})$.
Since $u_i$ is continuous, we obtain
$u_i(\underline{a}_0,a^*) \geq u_i(a_0^*,a'_i,a^*_{-i})$
by taking limits. This proves $a^* \in \mathcal{E}(f_{a^*})$ and thus $a^* \in \mathcal{E}^*$.
\end{IEEEproof}

Now we turn to the question of whether the best feasible performance, $\overline{v}_0$,
can be achieved with intervention. At an intervention equilibrium of the form $(f_{a^*}, a^*)$,
intervention exists only as a threat to deter deviation, and no intervention is exerted
as long as the users follow the target action profile. This contrasts with the imperfect
monitoring scenario considered in Proposition~\ref{thm:ineff}, where providing incentives requires that
intervention be used sometimes even when the users follow the target action profile, which results in a performance loss.
Thus, with perfect monitoring, it is possible for an intervention scheme to achieve the best feasible performance
as long as the intervention capability is sufficiently strong. This discussion
is formally stated below as a corollary of Proposition~\ref{prop:charie}.
Note that $\overline{v}_0 = \sup_{a \in A} u_0(\underline{a}_0,a)$, which is attained
when $A$ is compact.

\begin{corollary} \label{cor:eff}
If $a^o \in \arg \max_{a \in A} u_0(\underline{a}_0,a)$ and $u_i(\underline{a}_0,a^o)
\geq u_i(f_{a^o}(a_i,a_{-i}^o),a_i,a_{-i}^o)$ for all $a_i \in A_i$, for all $i \in \mathcal{N}$,
then $v_0^* = \overline{v}_0$.
\end{corollary}

Extreme intervention rules are useful to characterize sustainable action profiles
and intervention equilibria. However, they may not be desirable in practice.
For example, when a user chooses an action different from the target action
by mistake (i.e., trembling hands), an extreme intervention rule triggers the
most severe punishment for the user, which may result in a large performance loss. Thus, it is
of interest to investigate intervention rules that use weaker punishments than
extreme intervention rules do. To obtain concrete results, we assume that
$A_i = [\underline{a}_i, \overline{a}_i] \subset
\mathbb{R}$ with $\underline{a}_i < \overline{a}_i$ for all $i \in \mathcal{N}_0$
in the remainder of this subsection. Below we define another class of intervention rules.

\begin{definition}
$f_{\tilde{a},c}:A \rightarrow A_0$ is a \emph{(truncated) affine intervention
rule with target action profile $\tilde{a} \in A$ and intervention rate profile $c \in \mathbb{R}^N$} if
\begin{align*} 
f_{\tilde{a},c}(a) = \left[ c \cdot (a - \tilde{a}) + \underline{a}_0 \right]_{\underline{a}_0}^{\overline{a}_0},
\end{align*}
where $[x]_{\alpha}^{\beta} = \min \{ \max\{x,\alpha\}, \beta\}$.
\end{definition}

The following proposition constructs an affine intervention rule
to sustain an interior target action profile in the differentiable payoff case.

\begin{proposition} \label{prop:affine}
Let $a^* \in \mathcal{A}$ be an action profile such that $a_i^* \in (\underline{a}_i, \overline{a}_i)$ for all $i \in \mathcal{N}$.
Suppose that, for all $i \in \mathcal{N}$, $u_i$ is twice continuously differentiable and $u_i(a_0,a^*)$ is strictly decreasing in
$a_0$ on $[\underline{a}_0, \overline{a}_0]$.
Let
\begin{align} \label{eq:ci}
c_i^* = - \frac{\partial u_i(\underline{a}_0, a^*)/\partial a_i}{\partial u_i(\underline{a}_0, a^*)/\partial a_0}
\end{align}
for all $i \in \mathcal{N}$.\footnote{We define $\partial u_i(\underline{a}_0, a^*)/\partial a_0$ as the right partial derivative of
$u_i$ with respect to $a_0$ at $(\underline{a}_0, a^*)$.} Suppose that
\begin{align*} 
\frac{\partial^2 u_i}{\partial a_i^2} (\underline{a}_0, a_i, a_{-i}^*) \leq 0 \quad \text{for all $a_i \in (\underline{a}_i, \overline{a}_i)$}
\end{align*}
for all $i \in \mathcal{N}$ such that $c_i^* = 0$,
\begin{align*} 
\frac{\partial^2 u_i}{\partial a_i^2} (\underline{a}_0, a_i, a_{-i}^*) \leq 0 \quad \text{for all $a_i \in (\underline{a}_i, a_i^*)$},
\end{align*}
\begin{align*} 
\left( (c_i^*)^2 \frac{\partial^2 u_i}{\partial a_0^2}
+ 2 c_i^* \frac{\partial^2 u_i}{\partial a_i \partial a_0}
+ \frac{\partial^2 u_i}{\partial a_i^2} \right) \bigg|_{(a_0, a_i, a_{-i}) = (c_i^* (a_i - a_i^*) + \underline{a}_0, a_i, a_{-i}^*)} \leq 0 \nonumber \\
\quad \text{for all $a_i \in (a_i^*, \min \{\overline{a}_i, a_i^* + (\overline{a}_0 - \underline{a}_0)/c_i^* \})$, and}
\end{align*}
\begin{align*} 
\frac{\partial u_i}{\partial a_i} (\overline{a}_0, a_i, a_{-i}^*) \leq 0 \quad \text{for all $a_i
\in (a_i^* + (\overline{a}_0 - \underline{a}_0)/c_i^*, \overline{a}_i)$}
\end{align*}
for all $i \in \mathcal{N}$ such that $c_i^* > 0$, and
\begin{align*} 
\frac{\partial u_i}{\partial a_i} (\overline{a}_0, a_i, a_{-i}^*) \geq 0 \quad \text{for all $a_i \in
(\underline{a}_i, a_i^* + (\overline{a}_0 - \underline{a}_0)/c_i^*)$},
\end{align*}
\begin{align*} 
\left( (c_i^*)^2 \frac{\partial^2 u_i}{\partial a_0^2}
+ 2 c_i^* \frac{\partial^2 u_i}{\partial a_i \partial a_0}
+ \frac{\partial^2 u_i}{\partial a_i^2} \right) \bigg|_{(a_0, a_i, a_{-i}) = (c_i^* (a_i - a_i^*) + \underline{a}_0, a_i, a_{-i}^*)} \leq 0 \nonumber \\
\quad \text{for all $a_i \in (\max \{\overline{a}_i, a_i^* + (\overline{a}_0 - \underline{a}_0)/c_i^* \}, a_i^*)$, and}
\end{align*}
\begin{align*} 
\frac{\partial^2 u_i}{\partial a_i^2} (\underline{a}_0, a_i, a_{-i}^*) \leq 0 \quad \text{for all $a_i
\in (a_i^*, \overline{a}_i)$}
\end{align*}
for all $i \in \mathcal{N}$ such that $c_i^* < 0$.\footnote{We define $(\alpha, \beta) = \emptyset$ if $\alpha \geq \beta$.}
Then $f_{a^*,c^*}$ sustains $a^*$.
\end{proposition}

\begin{IEEEproof}
See the Appendix of \cite{ParkInfocom}.
\end{IEEEproof}

Note that $\partial u_i(\underline{a}_0, a^*)/\partial a_0 < 0$ for all $i \in \mathcal{N}$
since $u_i(a_0,a^*)$ is strictly decreasing in $a_0$. Thus, $c_i^*$, defined in \eqref{eq:ci}, has
the same sign as $\partial u_i(\underline{a}_0, a^*)/\partial a_i$. With $A_0 = [\underline{a}_0,
\overline{a}_0]$, the intervention action can be interpreted
as the intervention level, and at the target action profile $a^*$ the users receive higher payoffs as the intervention
level is smaller.
The affine intervention rule $f_{a^*,c^*}$, constructed in Proposition~\ref{prop:affine},
has the properties that the intervention device uses the minimum intervention level $\underline{a}_0$
when the users choose the target action profile $a^*$, i.e., $f_{a^*,c^*}(a^*) = \underline{a}_0$,
and that the intervention level increases in the rate of $|c_i^*|$ as
user $i$ deviates to the direction in which its payoff increases at $(\underline{a}_0, a^*)$.
The expression of $c_i^*$ in \eqref{eq:ci} has an intuitive explanation.
Since $c_i^*$ is proportional to $\partial u_i(\underline{a}_0, a^*)/\partial a_i$
and inversely proportional to $-\partial u_i(\underline{a}_0, a^*)/\partial a_0$,
a user faces a higher intervention rate as its incentive to deviate
from $(\underline{a}_0, a^*)$ is stronger and as a change in the intervention level has a smaller
impact on its payoff. The intervention level does not react to the action of user $i$
when $c_i^* = 0$, because user $i$ chooses $a_i^*$ in its self-interest
even when the intervention level is fixed at $\underline{a}_0$, provided that the other
users choose $a_{-i}^*$. Finally, we note that if $(f^*,a^*)$ is an
intervention equilibrium and $f_{a^*,c}$ sustains $a^*$ for some $c$,
then $(f_{a^*,c},a^*)$ is also an intervention equilibrium, since $f_{a^*,c}(a^*)
= \underline{a}_0$.

\subsection{Illustrative Example (Type-1 Intervention)} \label{sec:ex2}

As an illustrative example, we consider another resource sharing scenario
in a wireless network where $N \geq 2$ users and an
intervention device interfere with each other. In this example, the intervention
device engages in type-1 intervention, affecting the service quality
through its usage level.
The actions of the users and the intervention device are their usage levels,
and the action space is given by $A_i = [0, \overline{a}_i]$ for
all $i \in \mathcal{N}_0$. $\overline{a}_i$ denotes the maximum usage level of
user $i$, and $\overline{a}_0$ denotes that of the intervention device, which can be
considered as its intervention capability. We assume that $\overline{a}_i \geq q/2b$ for
all $i \in \mathcal{N}$, while imposing no restriction on $\overline{a}_0$.
The service quality is determined by the total usage level, $a_0 + \sum_{i=1}^N a_i$, following
the relationship
\begin{align*}
Q(a_0, a) = \left[q - b\left(a_0 + \sum_{i=1}^N a_i\right)\right]^+,
\end{align*}
where $q,b > 0$ and $[x]^+ = \max \{x, 0\}$. The payoff of
user $i \in \mathcal{N}$ is given by the product of the service quality and its
own usage level,
\begin{align} \label{eq:payoffexample}
u_i(a_0, a) = Q(a_0, a) a_i.
\end{align}
The payoff of the manager is given by the average payoff of the users,
\begin{align*}
u_0(a_0, a) = \frac{1}{N} \sum_{i=1}^N u_i(a_0, a).
\end{align*}
$u_i(a_0, a)$ is weakly decreasing in $a_0$ for all $a$, and thus we can consider
an extreme intervention rule that takes the value $\overline{a}_0$ whenever
a unilateral deviation occurs.

In this example, we have $\overline{v}_0 = q^2/4Nb$, which is achieved
when $a_0 = 0$ and $\sum_{i=1}^N a_i = q/2b$.
The symmetric action profile
that attains $\overline{v}_0$ is thus $(a_l,\ldots,a_l)$,
where $a_l \triangleq q/2Nb$. On the other hand, the best performance at the non-cooperative
equilibrium without intervention (i.e., when $a_0$ is held
fixed at 0) is given by $\tilde{v}_0 = q^2/(N+1)^2b$, which is attained at $(a_h,\ldots,a_h)$,
where $a_h \triangleq q/(N+1)b$. Note that $a_h > a_l$.
Hence, the goal of the manager
is to limit the usage levels of the users by using intervention as a threat.
In the following proposition, we investigate the best performance with intervention, $v_0^*$,
as we vary $\overline{a}_0$.

\begin{proposition} \label{prop:ex2}
(i) $v_0^* = \tilde{v}_0$ if and only if $\overline{a}_0 = 0$.\\
(ii) $v_0^* = \overline{v}_0$ if and only if $\overline{a}_0 \geq \overline{a}_0^{min} \triangleq (\sqrt{N} - 1)^2 q/ 2 N b$.\\
(iii) $v_0^*$ is strictly increasing with $\overline{a}_0$ on $[0, \overline{a}_0^{min}]$.
\end{proposition}

\begin{IEEEproof}
See Appendix~\ref{app:ex2}.
\end{IEEEproof}

\begin{figure}%
\centering
\subfloat[][$\overline{a}_0 = 0.1$]{%
\label{fig:a0vary-a}%
\hspace{-10mm}
\includegraphics[width=0.4\textwidth]{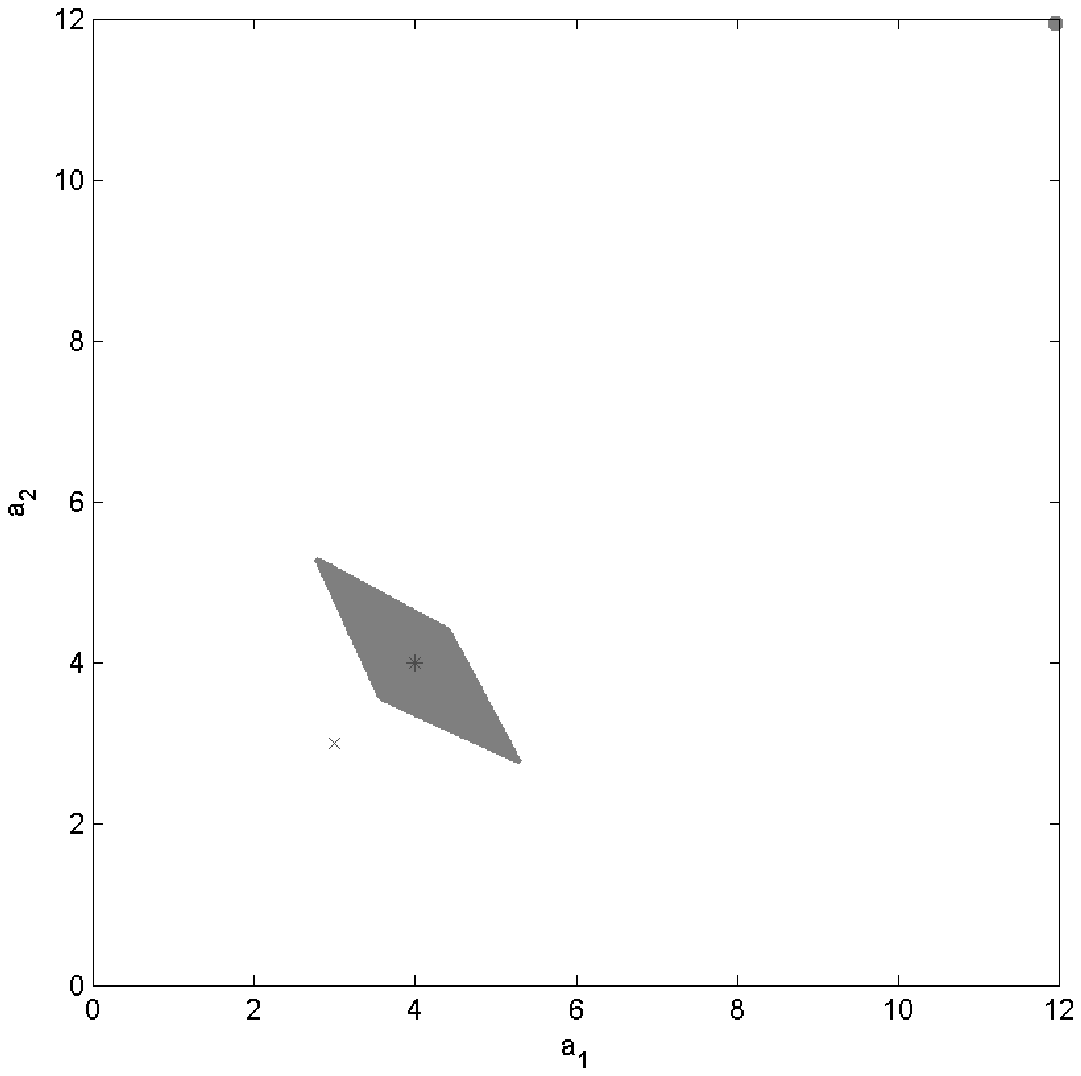}}%
\hspace{14mm}
\subfloat[][$\overline{a}_0 = 0.51$]{%
\label{fig:a0vary-b}%
\includegraphics[width=0.4\textwidth]{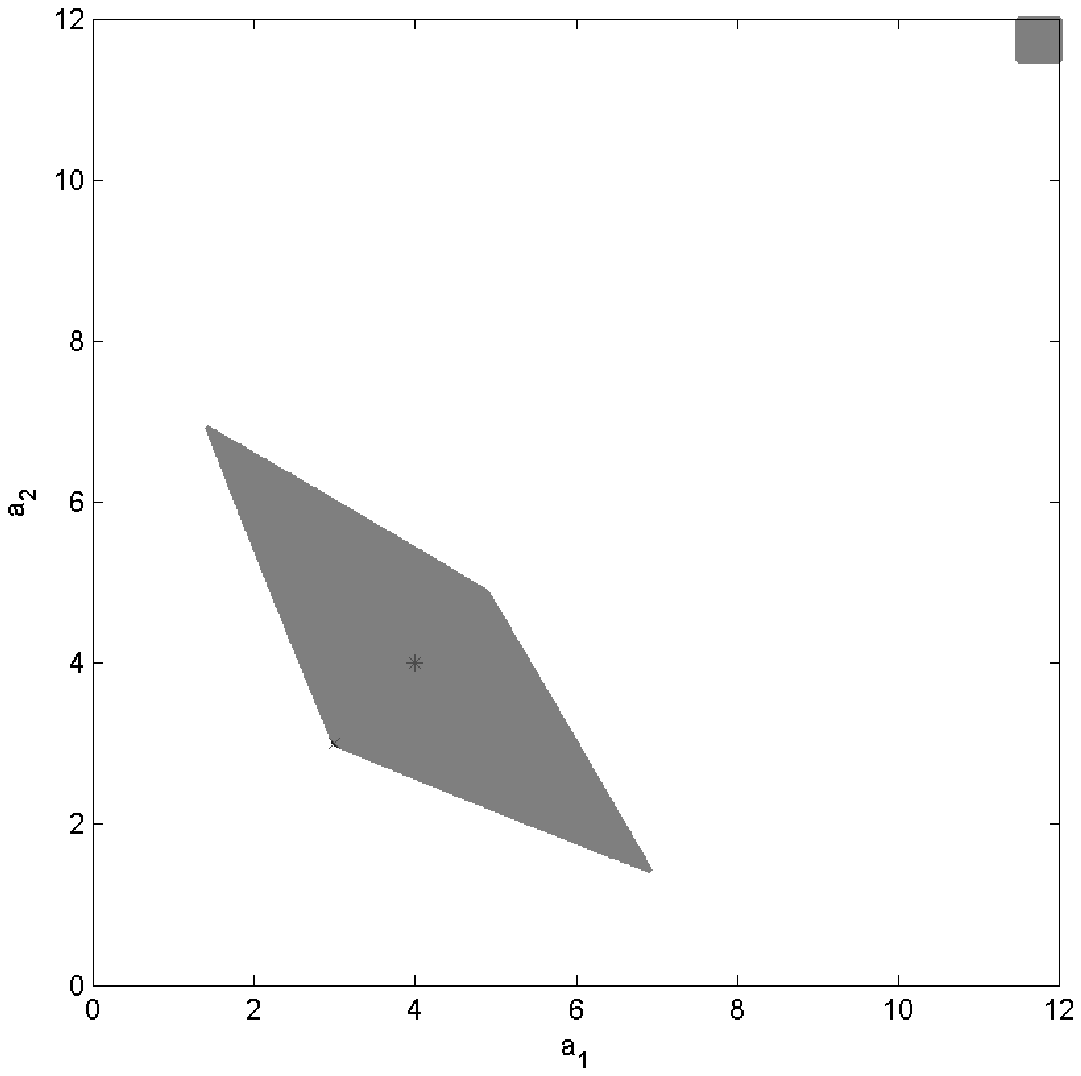}}\\
\vspace{5mm}
\subfloat[][$\overline{a}_0 = 5$]{%
\label{fig:a0vary-c}%
\hspace{-10mm}
\includegraphics[width=0.4\textwidth]{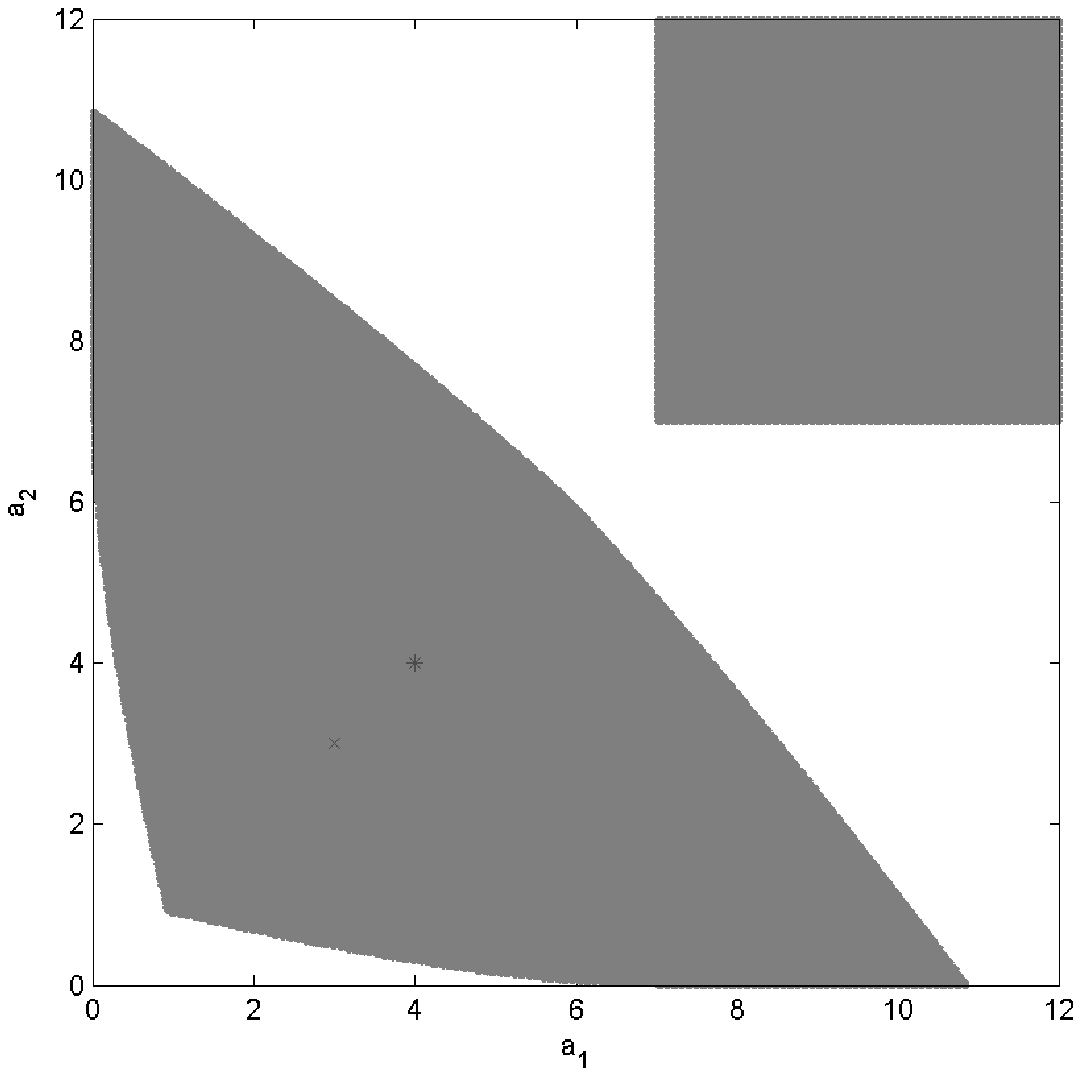}}%
\caption{Plot of $\mathcal{E}^*$ as dark regions for the different values of
$\overline{a}_0$ in the example in Section~\ref{sec:ex2}.}
\label{fig:a0vary}%
\end{figure}

Since $u_i$ is weakly decreasing in $a_0$, the set $\mathcal{E}^*$
is weakly expanding as the intervention capability $\overline{a}_0$
is larger. This implies that the performance with intervention
$v_0^*$ is weakly increasing with $\overline{a}_0$. Proposition~\ref{prop:ex2}
shows that the performance with intervention improves
as $\overline{a}_0$ increases, eventually reaching the best feasible
performance when $\overline{a}_0 \geq \overline{a}_0^{min}$.
Thus, $\overline{a}_0^{min}$ can be interpreted as the minimum intervention
capability for an intervention scheme to achieve the best feasible performance.
We can show that $\overline{a}_0^{min}$ is increasing and concave in $N$.
Fig.~\ref{fig:a0vary} plots the set $\mathcal{E}^* = \mathcal{E}(\mathcal{F})$ as
dark regions for the different values of $\overline{a}_0$ with parameters $N = 2$, $q=12$, $b=1$,
and $\overline{a}_1 = \overline{a}_2 = 12$. We can see that
$\mathcal{E}^*$ expands as $\overline{a}_0$ increases.
When $\overline{a}_0 = 0$, $\mathcal{E}^*$ has only two elements,
$(a_h,a_h)=(4, 4)$ and $(12, 12)$. When $\overline{a}_0 = 0.1$, there are more
action profiles in $\mathcal{E}^*$. However, the symmetric
social optimum $(a_l,a_l)=(3,3)$ does not belong to $\mathcal{E}^*$,
and Proposition~\ref{prop:charie} implies that the action profile
$(a_1,a_2)$ that minimizes $a_1 + a_2$ among those in $\mathcal{E}^*$ constitutes an intervention equilibrium.
When $\overline{a}_0 \geq (\sqrt{2} - 1)^2 q/ 4 b \approx 0.51$, the action
profiles in $\mathcal{E}^*$ that satisfy $a_1 + a_2 = 2a_l = 6$
constitute an intervention equilibrium, as all of them yield the
best feasible performance $\overline{v}_0$. When $\overline{a}_0 \geq q/b =12$,
the punishment from $\overline{a}_0$ is strong enough to make any
action profile sustainable, i.e., $\mathcal{E}^* = A$.

\begin{figure}%
\centering
\includegraphics[width=0.5\textwidth]{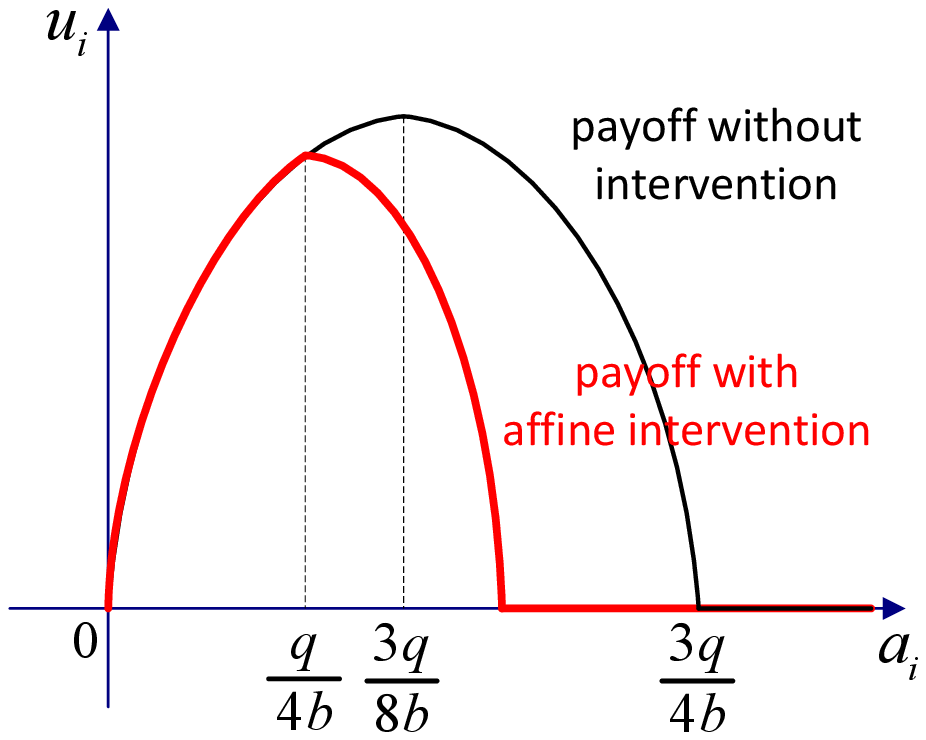}
\caption{Plot of $u_i$ against $a_i$ when the manager chooses
the affine intervention rule \eqref{eq:affineexample} and the other user
chooses $a_l$.} \label{fig:payoffex}
\end{figure}

Applying Proposition~\ref{prop:affine}, we can construct an affine intervention rule
that sustains an action profile $a^*$ such that $a_i^* \in
(0, \overline{a}_i)$ for all $i \in \mathcal{N}$ and $\sum_{i=1}^N a_i^* <q/b$, provided
that the maximum intervention level $\overline{a}_0$ is sufficiently large.
With the payoff functions in \eqref{eq:payoffexample}, the
expression of $c_i^*$ in \eqref{eq:ci} is given by
\begin{align*}
c_i^*(a^*) = \frac{q}{ba_i^*} - \frac{\sum_{j \neq i} a_j^*}{a_i^*} - 2,
\end{align*}
for all $i \in \mathcal{N}$. For example, the affine intervention rule with target
action profile $(a_l,\ldots,a_l)$ and the corresponding intervention rate profile $c^*(a_l,\ldots,a_l)$ is expressed as
\begin{align} \label{eq:affineexample}
f(a) = \left[ (N-1) \left(\sum_{i=1}^N a_i - \frac{q}{2b} \right) \right]_0^{\overline{a}_0}.
\end{align}
Fig.~\ref{fig:payoffex} considers $N = 2$ and plots the payoff of
user $i$ against its action $a_i$, provided that the manager
chooses the intervention rule in \eqref{eq:affineexample} and the
other user chooses $a_l$. It also assumes that
$\overline{a}_0$ is sufficiently large. Without intervention, the best
response of user $i$ to $a_l$ is $3q/8b$, which shows the
instability of the symmetric social optimum $(a_l,a_l)$. However,
when the intervention rule \eqref{eq:affineexample} is used, the
intervention device begins to intervene as user $i$ increases its
usage level from $a_l$. An increase in payoff due to the increased
usage level is more than offset by a decrease in payoff due to the
quality degradation from intervention. As a result, users
do not gain by a unilateral deviation from $(a_l,a_l)$ under the
intervention rule \eqref{eq:affineexample}.

\section{Comparison with Existing Approaches} \label{sec:compare}

The literature has studied various methods to improve non-cooperative outcomes.
One such method is to use contractual agreements. Contract theory is a field
of economics that studies how economic actors form contractual agreements, covering the
topics of incentives, information, and institutions \cite{bolton}.
Since intervention schemes aim to motivate users to take appropriate actions,
our work shares a theme as well as a formal framework with contract theory.
However, most works in contract theory deal with the principal-agent problem
using monetary payment as the incentive device (see, for example, \cite{holmstrom}).
In contrast, our work focuses on the problem of regulating selfish behavior
in resource sharing by using intervention within the system as the incentive device.

In game theory, correlated equilibrium is a solution concept that extends
Nash equilibrium and thus has the potential to improve Nash equilibrium.
A correlated equilibrium can be implemented by having a mediator who determines
an action profile following a correlated distribution and makes a confidential
recommendation to each player \cite{myerson}. In an intervention game, the manager
recommends a pure or mixed action profile to users but does not use a correlated
distribution to determine the target action profile. Another difference is that an intervention
scheme uses an external punishment device to prevent deviation, which is not present in the concept of correlated equilibrium.
We also note that, for the prisoner's dilemma game where there is a dominant strategy
for each player, the set of correlated equilibria coincides with that of Nash equilibria.
This suggests that correlated equilibrium is more useful for inducing coordination (see, for example,
\cite{altman}, which considers a multiple access network) than for achieving cooperation
in a prisoner's dilemma scenario, as considered in this paper.

Another method used in game theory to expand the set of Nash equilibria is repeated games.
In a repeated game, players monitor their behavior and
choose their actions based on past observations (see, for example, \cite{Cagalj} and \cite{buchegger}
for works that apply the idea of repeated games to wireless communications). Implementing an
incentive scheme based on a repeated game strategy requires
long-term relationship among interacting users, which may not exist
especially in mobile, cognitive, and vehicular networks.
Moreover, a repeated game strategy should be designed in accord with the self-interest
of players in order to ensure that they execute monitoring and punishment or reward in a planned way.
On the contrary, an intervention scheme uses an external device for monitoring and executing
punishment. Hence, it can provide incentives for a dynamically changing population,
and the manager can prescribe any feasible intervention rule according to his objective.

In the communications literature, Stackelberg games have been used to improve
Nash equilibrium (see, for example, \cite{ma} and \cite{korilis}). Stackelberg
games divide users into two groups, a leader and followers, and the leader takes an action
before the followers do. In intervention games, the manager is the leader while users are
followers, and the manager chooses an intervention rule, which is a contingent plan, instead of an action.
Thus, intervention games are more suitable than Stackelberg games when the leader is not a resource
user but a manager who regulates resource sharing by users.

Pricing schemes or taxation can also be used to induce individuals to
take socially desirable actions. Intervention affects the payoffs of users by directly influencing
their resource usage, whereas pricing does so by using an outside
instrument, money. Thus, intervention schemes can be implemented
more robustly in that users cannot avoid intervention as long as
they use resources. In order to achieve a desired outcome
through an incentive scheme, the manager needs to know the impact of the
incentive device on the payoffs of users. Since intervention
affects the payoffs of users through physical quantities associated
with resource usage (e.g., throughput, delay), which are easily
measurable, this information is easier to obtain when the manager
uses an intervention scheme than a pricing scheme, as discussed at the end of
Section~\ref{sec:ex1}.

Lastly, we discuss the difference between intervention
and mechanism design in the sense of \cite[Ch. 23]{mas}.
In a mechanism design problem, the designer aims to obtain the private
information of agents while he can control the social choice (e.g., a resource allocation).
On the contrary, in an intervention game, the manager aims to motivate
users to take appropriate actions while he has complete information
about users (i.e., no private information).

\section{Conclusion and Future Research} \label{sec:conclusion}

In this paper, we have developed a game-theoretic framework for the design and analysis
of intervention schemes, which are aimed to drive self-interested users
towards a system objective. Our results suggest that the manager can construct
an effective intervention scheme when he has an intervention device
with an accurate monitoring technology and a strong intervention capability.
We have illustrated our framework and results with simple resource sharing scenarios
in wireless communications. However, the application of intervention schemes
is not limited to the problems considered in this paper; our framework can
be applied to a much broader set of problems in communications, including power control
and flow control, as well as to various types of networks such as cognitive radio, vehicular
networks, peer-to-peer networks, and crowdsourcing websites. Exploring the role
of intervention in various specific scenarios is left for future research.
Another direction of future research is to combine intervention with other
game-theoretic concepts. First, we can introduce intervention in repeated games,
where users and the intervention device choose their actions depending on
their past observations. We can also allow the intervention manager to use a
correlated distribution, as in correlated equilibrium, when he determines
the target action profile. Intervention can then be exerted when a user deviates
from the recommended action. We can use the idea of mechanism design
to deal with a scenario where the intervention manager has incomplete information about
users. In such a scenario, the manager first obtains reports from users and then
chooses an intervention rule depending on the reports. Finally, intervention
can be used in the context of bargaining games, where the set of feasible
payoffs in a bargaining game is obtained from sustainable action profiles.

\appendices

\section{Proof of Proposition~\ref{prop:ex1}} \label{app:ex1}

\begin{IEEEproof}
First, note that $w_0(0) > \lim_{\alpha \rightarrow 0^+} w_0(\alpha)$.
Suppose that $p a_L - q a_H < 0$ and $q a_L - r a_H < 0$. Then $w_0(\alpha) = 0$ for all $\alpha \in (0,1]$,
and thus $v_0^* = w_0(0) = \tilde{v}_0$. This covers condition (a) in Proposition~\ref{prop:ex1}.
Now suppose that at least one of the two inequalities $p a_L - q a_H \geq 0$ and $q a_L - r a_H \geq 0$
holds. We consider three cases.

\emph{Case 1:}  $p a_L - q a_H = q a_L - r a_H$.

In this case, $\alpha (p a_L - q a_H) + (1-\alpha) (q a_L - r a_H) \geq 0$ is
satisfied for all $\alpha \in (0,1]$, and $w_0(\alpha)$ is increasing on $(0,1]$. Thus, we obtain
$v_0^* = \max \{w_0(0), w_0(1) \}$.

\emph{Case 2:}  $p a_L - q a_H > q a_L - r a_H$.

$\alpha (p a_L - q a_H) + (1-\alpha) (q a_L - r a_H) \geq 0$
if and only if
\begin{align} \label{eq:case2al}
\alpha \geq \frac{- (q a_L - r a_H)}{(p a_L - q a_H) - (q a_L - r a_H)},
\end{align}
where the right-hand side of \eqref{eq:case2al} is smaller than 1.
Also, $p a_L - q a_H > q a_L - r a_H$ implies $p - q > q - r$.
We can show that the sign of the first derivative of $w_0$ at any $\alpha \in (0,1)$ is equal to that of
$(p-q)(1-r) - (q-r)(1-q)$, which is positive. Hence, we have $v_0^* = \max \{w_0(0), w_0(1) \}$.
Combining Cases 1 and 2 covers conditions (c) and (d).

\emph{Case 3:}  $p a_L - q a_H < q a_L - r a_H$.

$\alpha (p a_L - q a_H) + (1-\alpha) (q a_L - r a_H) \geq 0$
if and only if $\alpha \leq \overline{\alpha}$.
Also, the sign of the first derivative of $w_0$ is equal to that of
$(p-q)(1-r) - (q-r)(1-q)$.

\emph{Case 3-1:} $0 \leq p a_L - q a_H < q a_L - r a_H$.

We have $\overline{\alpha} \geq 1$.
Thus, $w_0$ is increasing on $(0,1]$ if $(p-q)(1-r) - (q-r)(1-q) > 0$ and non-increasing if $(p-q)(1-r) - (q-r)(1-q) \leq 0$.

\emph{Case 3-2:} $p a_L - q a_H < 0 \leq q a_L - r a_H$.

We have $\overline{\alpha} < 1$.
Thus, $w_0$ is increasing on $(0,\overline{\alpha}]$ if $(p-q)(1-r) - (q-r)(1-q) > 0$ and non-increasing if $(p-q)(1-r) - (q-r)(1-q) \leq 0$.

These results cover conditions (b), (e), and (f).
\end{IEEEproof}

\section{Proof of Proposition~\ref{prop:ex2}} \label{app:ex2}

\begin{IEEEproof}
(Sketch) Note that $u_0(0,a)$ depends on $a$ only through $\sum_{i=1}^N a_i$. $u_0(0,a)$
is increasing in $\sum_{i=1}^N a_i$ for $0 \leq \sum_{i=1}^N a_i \leq q/2b$,
reaches the maximum at $\sum_{i=1}^N a_i = q/2b$, is decreasing in $\sum_{i=1}^N a_i$
for $q/2b \leq \sum_{i=1}^N a_i \leq q/b$, and remains at zero for $\sum_{i=1}^N a_i \geq q/b$.

(i) If $\overline{a}_0 = 0$, then $v_0^* = \tilde{v}_0$ by definition.
To show the converse, suppose that $\overline{a}_0 > 0$. Since the payoff function is
continuous, we can show that $(a_h - \epsilon, \ldots, a_h - \epsilon)$ is sustainable for sufficiently small
$\epsilon > 0$, which yields $v_0^* > \tilde{v}_0$.

(ii) We have $v_0^* = \overline{v}_0$ if and only if there exists a sustainable
action profile $a$ such that $\sum_{i=1}^N a_i = q/2b$. Given $\sum_{i=1}^N a_i = q/2b$,
the incentive for user $i$ to deviate is stronger as $a_i$ is smaller. Hence,
it suffices to check whether the symmetric action profile $(a_l,\ldots,a_l)$ is
sustainable. By Lemma~\ref{lem:ext}, $(a_l,\ldots,a_l)$ is sustainable if and only if
\begin{align*}
\max_{a_i \in [0, \overline{a}_i]} [q - b(\overline{a}_0 + (N-1)a_l + a_i)]^+ a_i \leq q^2/4Nb,
\end{align*}
which is equivalent to $\overline{a}_0 \geq (\sqrt{N} - 1)^2 q/ 2 N b$.

(iii) Choose $\overline{a}_0, \overline{a}'_0 \in [0, \overline{a}_0^{min}]$ with
$\overline{a}_0 < \overline{a}'_0$. Let $v_0^*$ and $(v_0^*)'$ be the corresponding
performances with intervention. Since $0 \leq \overline{a}_0 < \overline{a}_0^{min}$,
there exists an action profile $a$ that attains $v_0^*$ with intervention capability $\overline{a}_0$ and satisfies
$q/2b < \sum_{i=1}^N a_i \leq Nq/(N+1)b$. We can show that $(a_1 - \epsilon, \ldots, a_N - \epsilon)$
can be sustained with $\overline{a}'_0$ for sufficiently small $\epsilon > 0$,
which implies $(v_0^*)' > v_0^*$.
\end{IEEEproof}

\newpage

\begin{table}
\caption{Payoff matrix of the game $\Gamma_{\tilde{f}}$ in the illustrative example in Section~\ref{sec:ex1}.}
\centering
\begin{tabular}{c|c|c|}
& $a_L$ & $a_H$ \\ \hline
$a_L$ & $y_p a_L$, $y_p a_L$ & $y_q a_L$, $y_q a_H$ \\ \hline
$a_H$ & $y_q a_H$, $y_q a_L$ & $y_r a_H$, $y_r a_H$ \\ \hline
\end{tabular}
\label{table:pd}
\end{table}

\begin{figure}%
\centering
\subfloat[][]{%
\label{fig:w0-a}%
\includegraphics[width=0.5\textwidth]{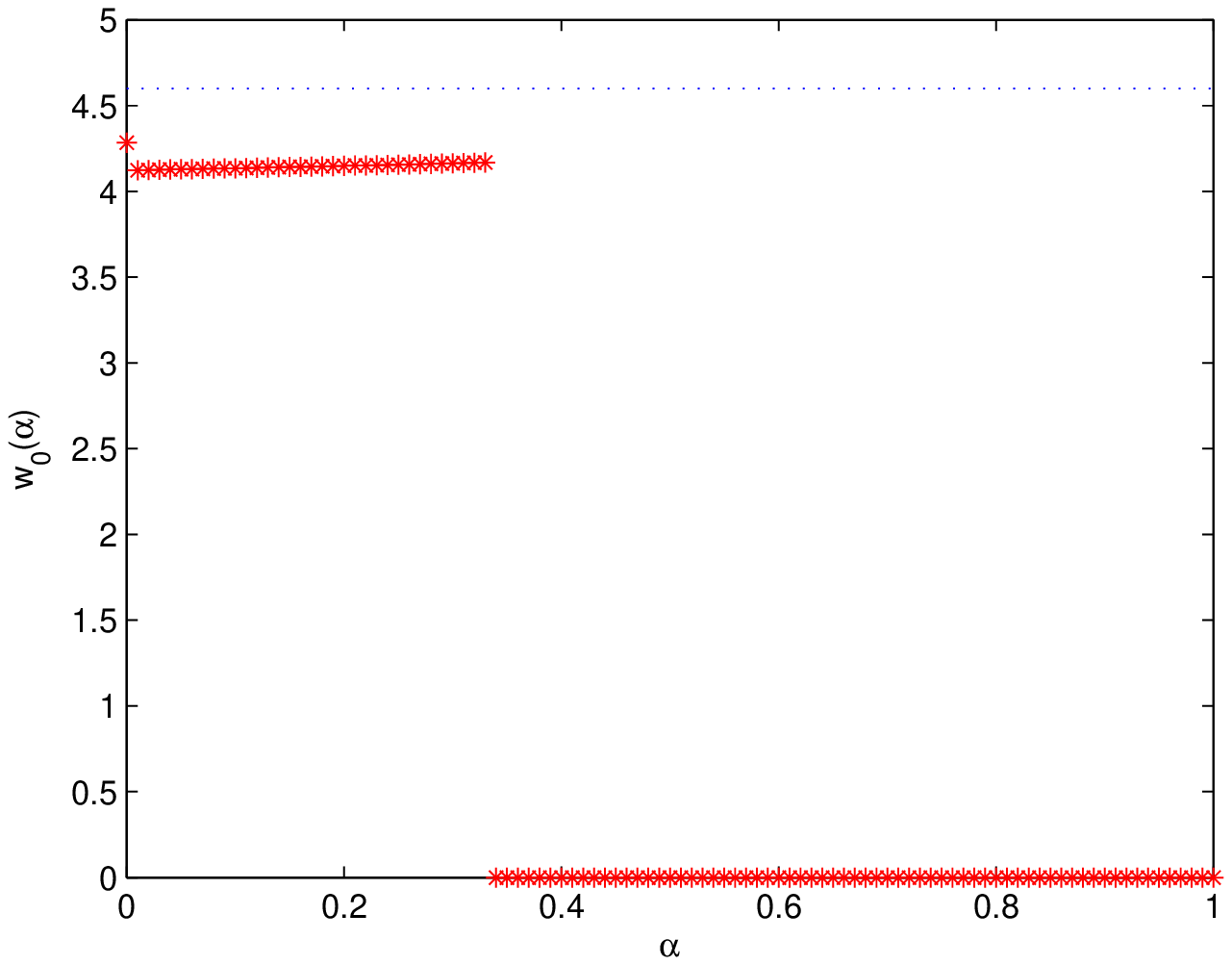}}%
\subfloat[][]{%
\label{fig:w0-b}%
\includegraphics[width=0.5\textwidth]{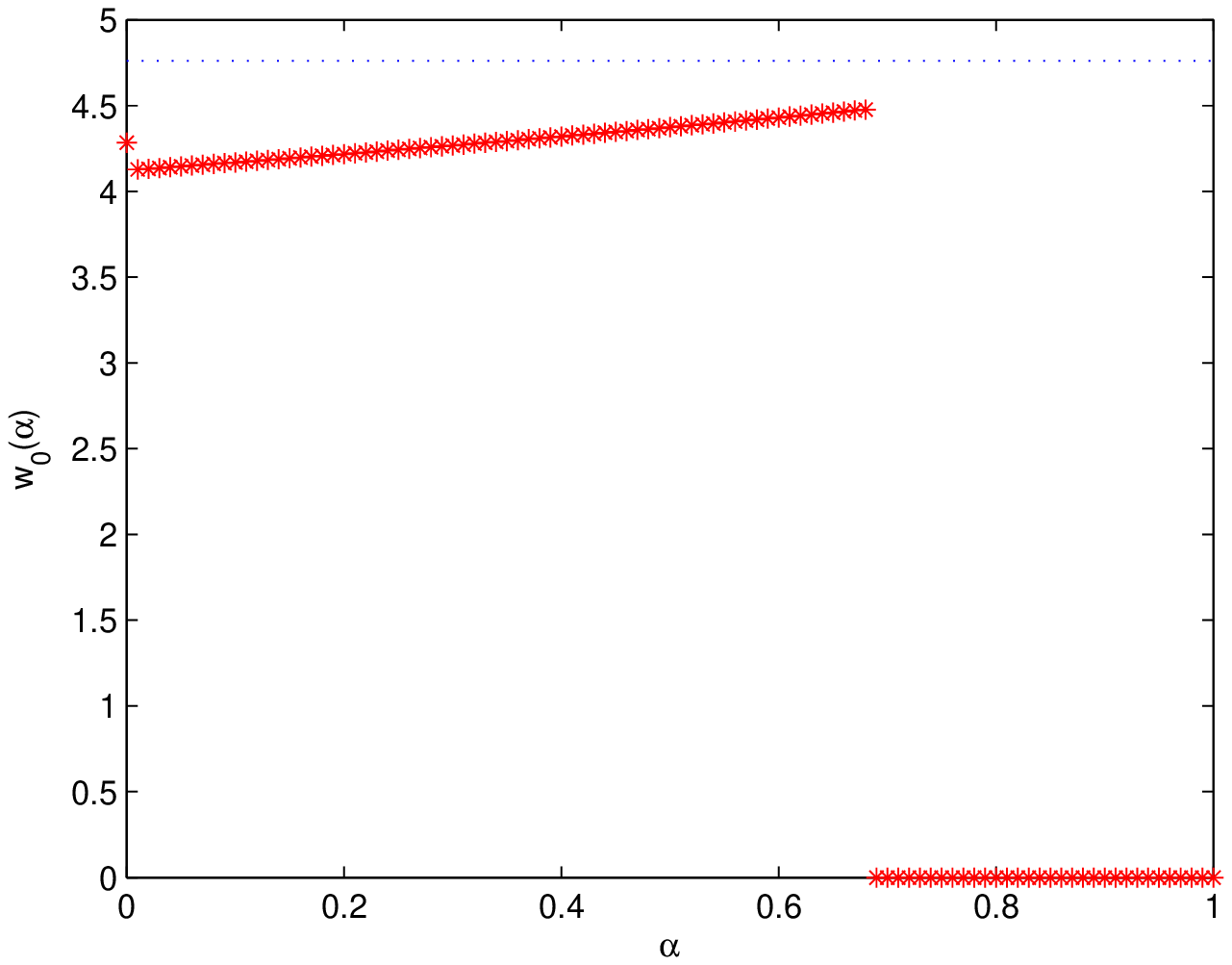}}\\
\subfloat[][]{%
\label{fig:w0-c}%
\includegraphics[width=0.5\textwidth]{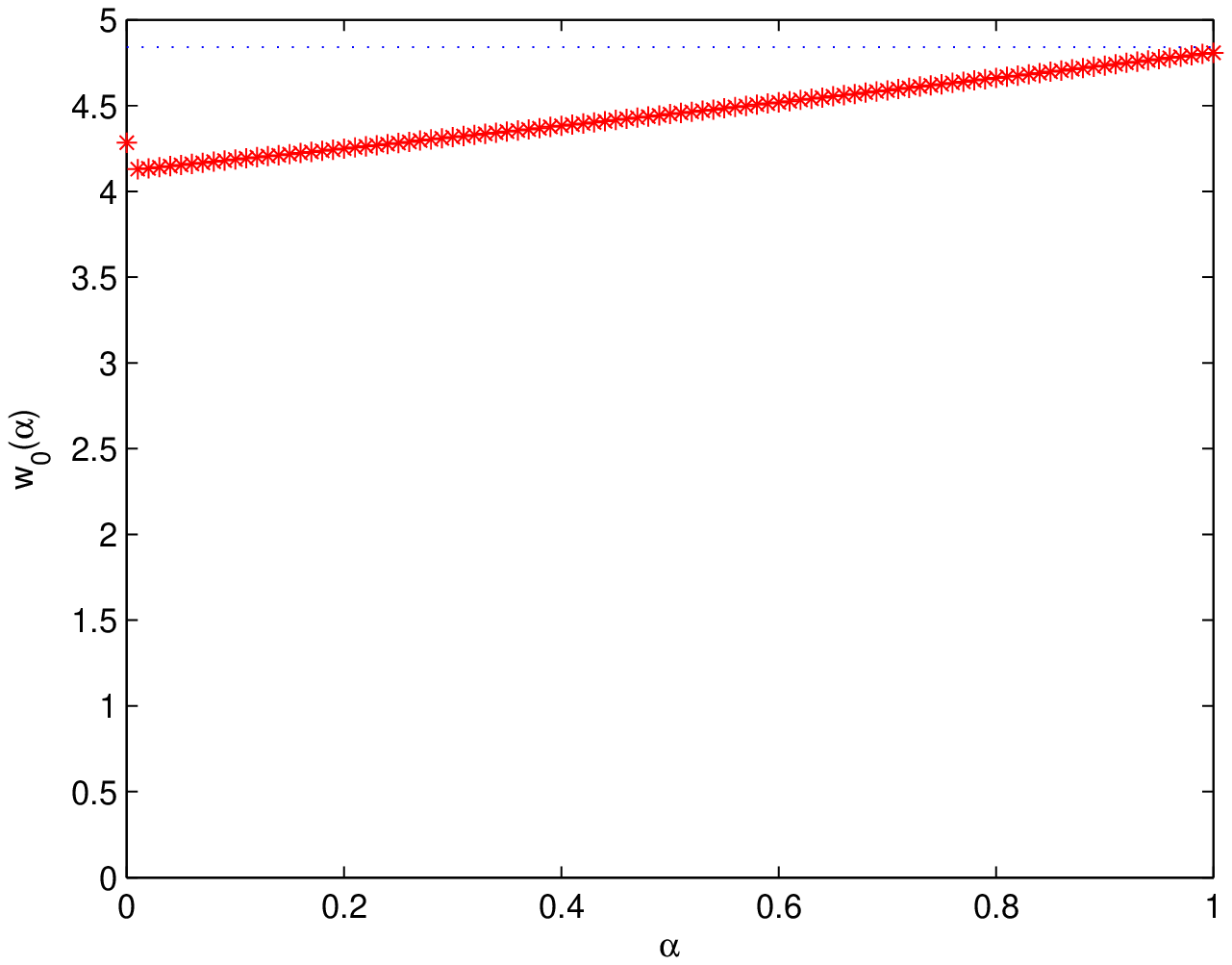}}%
\caption{The graph of the function $w_0(\alpha)$ defined in \eqref{eq:w0alpha}: \subref{fig:w0-a}
$v_0^* = \tilde{v}_0$ ($p = 0.9$), \subref{fig:w0-b} $v_0^* = w_0(\overline{\alpha})$ ($p = 0.94$), and
\subref{fig:w0-c} $v_0^* = w_0(1)$ ($p = 0.96$). (The dotted lines display $\overline{v}_0 = y_p a_L$.)}
\label{fig:w0}%
\end{figure}

\end{document}